\documentclass[pre,twocolumn]{revtex4-1}
\usepackage{graphicx}
\usepackage{bm}
\usepackage{epstopdf}
\usepackage{amsmath}
\usepackage{amssymb}

\newenvironment{eqn}{\begin{eqnarray*}}{\end{eqnarray*}}
\newenvironment{eqnn}{\begin{eqnarray}}{\end{eqnarray}}

\newcommand{\refeq}[1]{(\ref{#1})}

\newcommand{\dpart}[2]{\frac{\partial #1}{\partial{#2}}}
\newcommand{\dfunc}[2]{\frac{\delta #1}{\delta #2}}
\newcommand{\dtot}[2]{\frac{d#1}{d#2}}
\newcommand{\D}{\mathcal{D}}

\newcommand{\R}{\mathcal{R}}
\newcommand{\C}{\mathcal{C}}

\newcommand{\s}{\sigma}
\newcommand{\bs}{\boldsymbol{\sigma}}

\newcommand{\sh}{\hat{\sigma}}
\newcommand{\bsh}{\boldsymbol{\hat{\sigma}}}
\renewcommand{\d}{\delta}
\renewcommand{\l}{\lambda}
\newcommand{\bt}{\boldsymbol{\tau}}
\renewcommand{\th}{\hat{\tau}}
\newcommand{\bth}{\boldsymbol{\hat{\tau}}}
\newcommand{\x}{\xi}

\newcommand{\z}{\zeta}
\newcommand{\pmexp}[1]{#1_{i_{1}< i_{2} < \dots < i_{p}}}
\newcommand{\pexp}[1]{#1_{i_{1} i_{2} \dots i_{p}}}

\newcommand{\thav}[1]{\left<#1\right>}
\newcommand{\dav}[1]{\overline{#1}}

\newcommand{\bh}{\bm{h}}
\newcommand{\bl}{\bm{l}}
\newcommand{\bhh}{\bm{\hat{h}}}
\newcommand{\bhl}{\bm{\hat{l}}}

\newcommand{\etal}{\emph{et al.}\ }

\begin{document}

\title{Dynamical arrest with zero complexity: the unusual behavior of the
  spherical Blume Emery Griffiths disordered model}

\author{Corrado Rainone$^{1,2}$\email{Corrado.Rainone@roma1.infn.it}, Ulisse Ferrari$^{3}$, Matteo Paoluzzi$^1$ and Luca Leuzzi$^{4,1}$
  } \affiliation{
  $^1$ Dipartimento di Fisica,
  Sapienza Universit\`a di Roma, Piazzale A. Moro 2, I-00185 Roma,
  Italy \\
  $^2$ LPT, Ecole Normale Sup\'erieure, CNRS UMR 8549, 24 Rue
  Lhomond, 75005 Paris, France\\
  $^3$ Institut de la Vision, Sorbonne Universit\'es, UPMC, INSERM U968, CNRS UMR 7210, Paris, F-75012, France\\
   $^4$ NANOTEC-CNR, Soft and Living Matter Lab. Rome, c/o Dept. Physics, Sapienza
  Universit\`a di Roma, Piazzale A. Moro 2, I-00185 Roma, Italy 
  }

\date{\today}

\begin{abstract}
The short- and long-time dynamics of model systems undergoing a glass
transition with apparent inversion of Kauzmann and dynamical arrest
glass transition lines is investigated. These models belong to the
class of the spherical mean-field approximation of a spin-$1$ model
with $p$-body quenched disordered interaction, with $p>2$, termed
spherical Blume-Emery-Griffiths models. Depending on temperature and
chemical potential the system is found in a paramagnetic or in a
glassy phase and the transition between these phases can be of a
different nature. In specific regions of the phase diagram coexistence
of low density and high density paramagnets can occur, as well as the
coexistence of spin-glass and paramagnetic phases.  The exact static
solution for the glassy phase is known to be obtained by the one-step
replica symmetry breaking ansatz. Different scenarios arise for both
the dynamic and the thermodynamic transitions. These include: (i) the
usual random first- order transition (Kauzmann-like) for mean-field
glasses preceded by a dynamic transition, (ii) a thermodynamic
first-order transition with phase coexistence and latent heat and
(iii) a regime of apparent inversion of static transition line and
dynamic transition lines, the latter defined as a non-zero complexity
line.  The latter inversion, though, turns out to be preceded by a
novel dynamical arrest line at higher temperature.  Crossover between
different regimes is analyzed by solving mode coupling theory
equations throughout the space of external thermodynamic parameters
and the relationship with the underlying statics is discussed.

\end{abstract}

\maketitle

\section{Introduction\label{sec:intro}}
In the present work we investigate the dynamic properties
of a glassy system in which, under certain external conditions, both
glass and fluid can coexists, yielding different scenarios for
dynamical arrest and for the fluid-glass transition.  These properties
can be studied in statistical mechanical models with bosonic spin-$1$
variables, where the holes $s = 0$ play the role of inactive states,
that is, the so-called Blume-Capel \cite{Blume66,Capel66} or
Blume-Emery Griffiths (BEG) \cite{Blume71} models. In these models the
fluid phase corresponds to a paramagnet and the solid phase is either a
ferromagnet (no or weak disorder)
\cite{Blume66,Capel66,Blume71,Schupper04,Schupper05} or a spin glass
(strong disorder)
\cite{Ghatak77,Lage82,Mottishaw85,Sellitto97,Schreiber99,Crisanti02,Crisanti04d}.
In the present work we consider an extension to $p$-spin interacting
systems with spin-$1$, to $p>2$ and continuous (spherical) variables
\cite{Ferrari11} to better represent continuous density fluctuations,
alike to liquid-like compounds.  

In the presence of quenched disorder
the random BEG model with pairwise ($p=2$), as well as its spherical
counterpart, is known to display both a continuous
paramagnet/spin-glass phase transition and a first-order one (first
order in the thermodynamic sense, i.e. with latent heat and a region
of phase coexistence). Furthermore, {\em melting upon cooling}
\cite{Rastogi99, Greer00, vanRuth04,Plazanet04, Tombari05, Ferrari07,
  Angelini09} can occur, with a spin glass at high $T$ and a
paramagnet at low $T$. These properties have been observed in the
mean-field approximation, where the self-consistent solution for the
spin-glass phase is computed in the full replica symmetry breaking
(RSB) Parisi ansatz \cite{Crisanti05a} and on the cubic
3D lattice with nearest-neighbor couplings \cite{Paoluzzi10,Leuzzi11}. 
The frustrated BEG model has been
studied by means of numerical renormalization group techniques, as
well, with  results depending on the underlying lattice and the 
renormalization technique
 adopted \cite{Ozcelik08,Antenucci14a,Antenucci14b}.

Mean-field spin-glass models with Ising \cite{Gardner85}, soft
\cite{Kirkpatrick87a, Kirkpatrick87b} or spherical \cite{Crisanti92,Crisanti03b}
spins with more than two-spin interactions, called $p$-spin models,
are known to yield the so-called random first-order transition, i.e.,
a phase transition across which no internal energy discontinuity
occurs but the order parameter (the Edwards-Anderson overlap $q_{EA}$)
jumps from zero to a finite value. Their glassy phase is described by
an ansatz with one RSB \cite{Parisi79}.  In a cooling procedure, the
thermodynamic transition is preceded by a dynamic transition due to
the onset of a very large number of metastable states separated by
high barriers \cite{Leuzzi07}.  ``Very large'' means that the number
of states $\mathcal N$ grows exponentially with the size $N$ of the
system: $\mathcal N \sim \exp (\Sigma N)$ where the coefficient $\Sigma$
is the configurational entropy, also called {\em complexity} in the
framework of spin-glass systems (see, e.g, Refs. \cite{Mueller06,Crisanti04} and references therein). 
``High barriers'' means that the free
energy difference between a local minimum in the free energy
functional of the configurational space (also called free energy
landscape) and a nearby maximum (or saddle) grows with $N$. The
phenomenology of the $p$-spin spin-glass systems is, in many respects,
very similar to the one of structural glasses. These models are,
therefore, sometimes called mean-field glasses.  The occurrence of
non-zero $\Sigma$ is a fundamental property both in mean-field systems
\cite{Kirkpatrick87a, Kirkpatrick87b,Crisanti95} and outside the range
of validity of mean-field theory, e.g. in computer glass models
\cite{Sciortino05,Binder05}, or, indirectly, by measuring the excess
entropy of glasses in experiments, see, e.g., Ref. \cite{Leuzzi07} and
references therein.  The barriers' height turns out to diverge in the
thermodynamic limit in the mean-field approximation, this being an
artifact of mean-field glasses.  The thus induced dynamic transition
corresponds to the transition predicted by another mean-field theory
for the dynamics of supercooled liquids: the mode coupling theory
\cite{Goetze09}. The thermodynamic transition occurring at a lower
temperature is, instead, the mean-field equivalent of the so-called
Kauzmann transition in glasses, also known as the {\em ideal glass}
transition \cite{Leuzzi07}.  This was initially predicted by Gibbs and
Di Marzio \cite{Gibbs58} and its occurrence in real strutural glasses
is still object of an ongoing debate \cite{Hecksher08,Eckmann08,
  Tanaka03, Martinez-Garcia14}.

We are going to investigate the complex dynamic properties consequent
to the combination of a kind of interaction inducing structural glass
behavior and the presence of hole states (aka, spin state $s=0$)
inducing phase coexistence. The latter element is, possibly,
responsible for melting upon cooling \cite{Crisanti05a,Paoluzzi10}.
The first of such models was brought about by Sellitto in the pairwise
random orthogonal model with spin-$1$ variables \cite{Sellitto06}. In
the present dynamic work we rather consider the multi-body interaction
model of Ref. \cite{Ferrari11}, where both high temperature coexistence of high- 
and low-density paramagnetic phases, and low temperature coexistence of (low-density)
 paramagnetic and spin-glass phases are displayed.

\section{Model\label{sec:statics}}
The model we consider is a spherical Blume-Capel
\cite{Blume66,Capel66} model with $p$-body disordered interactions. Our
starting point is the model Hamiltonian
\begin{equation}
H = -\pmexp{\sum}\pexp{J}s_{i_{1}}s_{i_{2}}\dots s_{i_{p}} + D\sum_{i}s_{i}^{2},
\label{hbcpspin}
\end{equation}
where the variables $s_i$ are \emph{bosonic spins} (i.e.,
$s_i=1,0,-1$), and the couplings are independent quenched random
variables distributed with a Gaussian probability density,
$$
P(\pexp{J}) = \sqrt{\frac{N^{p-1}}{\pi J^2p!}}\exp\left[-\frac{\pexp{J^{2}}N^{p-1}}{p!}\right];
$$ the external parameter $D$ is called ``crystal field'' in
literature, and it essentially plays the role of a chemical
potential. Because of the bosonic spins, we cannot define the continous spin
approximation of the model with spherical constraint 
in the usual way \cite{Berlin52}. We must first rewrite the Hamiltonian as an
Ising-spin problem on a lattice-gas
\begin{equation}
\begin{split}
H = &-\pmexp{\sum}\pexp{J}n_{i_{1}}\s_{i_{1}}n_{i_{2}}\s_{i_{2}}\dots n_{i_{p}}\s_{i_{p}}\\ 
&+ (D-T\log2)\sum_{i}n_{i},
\end{split}
\end{equation}
where the $-T\log2$ term is necessary to keep the ratio of filled-in to empty sites identical to
the one of the original Hamiltonian, see \cite{Ferrari12,Griffiths67} for details. We then introduce the variable\cite{Caiazzo02}
\begin{equation}
\tau_{i} = \s_{i}(2n_{i}-1) = \pm 1.
\label{trasf}
\end{equation}
This way, the model Hamiltonian assumes the form
\begin{equation}
\begin{split}
H = &-\frac{1}{2^{p}}\pmexp{\sum}\pexp{J}(\s_{i_{1}}+\tau_{i_{1}})\dots(\s_{i_{p}}+\tau_{i_{p}})\\ 
&+ (D-T\log2) \sum_{i=1}^{N}\frac{\s_{i}\tau_{i}+1}{2},
\label{H}
\end{split}
\end{equation}
where all the degrees of freedom, $\sigma$'s and $\tau$'s are now Ising spins. 
A continuous spin model can then be constructed by imposing
 two independent spherical constraints
\begin{eqnn}
\sum_{i=1}^{N}\s_{i}^{2} = N, &\qquad & \sum_{i=1}^{N} \tau_{i}^{2} = N.
\label{costr}
\end{eqnn}
Programma/Pspinbosonic/Test/
The thermodynamic properties of the model we just defined are
thoroughly studied, applying 1RSB theory, in reference
\cite{Ferrari11}. For the sake of brevity, henceforth we shall
refer to Ref. \cite{Ferrari11} as the {\em static} study
(that is, replica theory-based), in contrast with the {\em dynamic}
study  that constitutes the
principal aim and subject of this paper.
Before deriving and solving
the dynamical equations of the model, it is suitable to briefly
summarize the static results of \cite{Ferrari11}, with particular
emphasis on the aspects that will be most relevant for the dynamical
study that we are going to report.

\subsection{The static phase diagram \label{sec:newstatics}}
The 1RSB free energy for the model is
\begin{equation}
\begin{split}
\beta F = &(\beta D - \log 2)d - \frac{\beta^{2}}{4}(d^{p} + (m-1)q_{1}^{p})-\log 2\\
&-\frac{1}{2}\left( \log(1-d) + \frac{m-1}{m}\log\eta_{0} + \frac{1}{m} \log\eta_{1}\right),
\end{split}
\label{eq:freeenergy}
\end{equation}
with the definitions
\begin{subequations}
\begin{eqnn}
\eta_{0} &\equiv& d-q_{1},\\
\eta_{1} &\equiv& d+(m-1)q_{1} - mq_{0},  
\end{eqnn}
\end{subequations}
where the parameter $d$ is the ratio of filled-in to empty sites (that
is, the density of the system), and the $q_0$ and $q_1$ are
respectively the mutual overlap and self-overlap, as usual in a 1RSB
Ansatz. The extremization of the \refeq{eq:freeenergy} whith respect
to $q_1$ and $d$ yields the saddle-point equations
\begin{subequations}
\begin{eqnn}
\frac{p\beta^{2}}{2}q_{1}^{p-1} &=& \frac{q_{1}}{\eta_{0}\eta_{1}},\label{eqsella1}\\
\frac{p\beta^{2}}{2}d^{p-1} -\frac{p\beta^{2}}{2}q_{1}^{p-1} &=& \frac{\eta_{0}-\theta}{\theta\eta_{0}} + 2(\beta D - \log2).\label{eqsella2}
\end{eqnn}
\end{subequations}
The static lines of the phase diagram, in absence of an external magnetic field, can be determined by setting $q_0=0$ and studying the system given by the equations \refeq{eqsella1}, \refeq{eqsella2}, and the saddle-point condition for the 1RSB parameter $m$
\begin{equation}
z(y) = \frac{2}{p}, \label{eqsella3}
\end{equation}
where $z(y)$ is the Crisanti-Sommers function\cite{Crisanti92}
\begin{equation}
z(y) \equiv -2y\frac{1-y + \log y}{(1-y)^{2}}, 
\qquad y \equiv \frac{\eta_{0}}{\eta_{1}}.
\end{equation}
The dynamical arrest lines can be, as well, identified using replica
theory, by solving the saddle-point equations for $m=1$, and by
imposing the following marginality condition for the solution in the
$d,q_1$ parameter space:
\begin{equation}
\left|
\begin{array}{cc}
\frac{\partial^{2}F}{\partial q_{1}^2} & 
\frac{\partial^{2}F}{\partial q_{1}\partial d} \\
& {\phantom .}\\
\frac{\partial^{2}F}{\partial d\partial q_{1}} & 
\frac{\partial^{2}F}{\partial d^2}
\end{array}
\right|
 = 0.
\end{equation}
This corresponds to looking for a spinodal point in the free energy as
a function of $d$ and $q_1$.

\begin{figure}[t!]
\begin{center}
\includegraphics[width = 0.99\columnwidth]{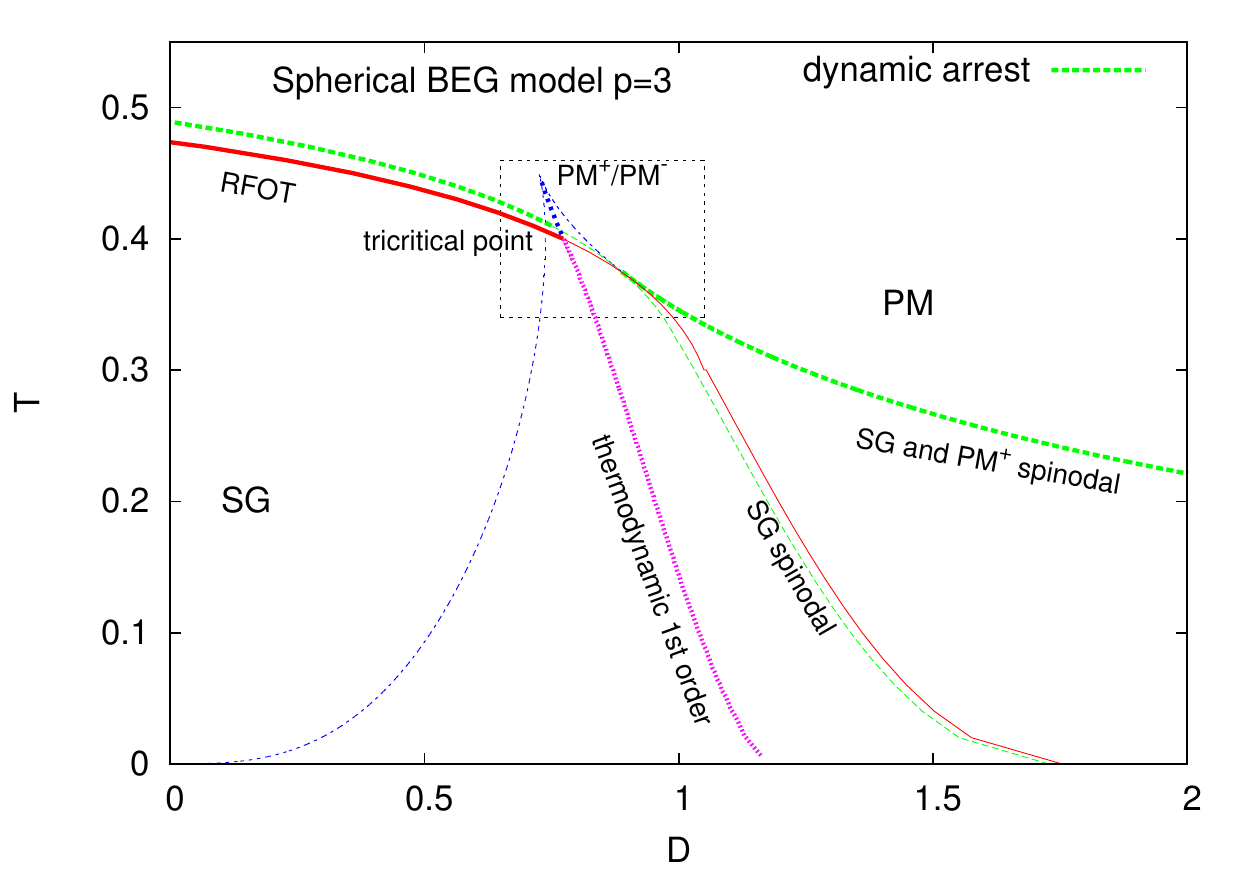}
\includegraphics[width = 0.99\columnwidth]{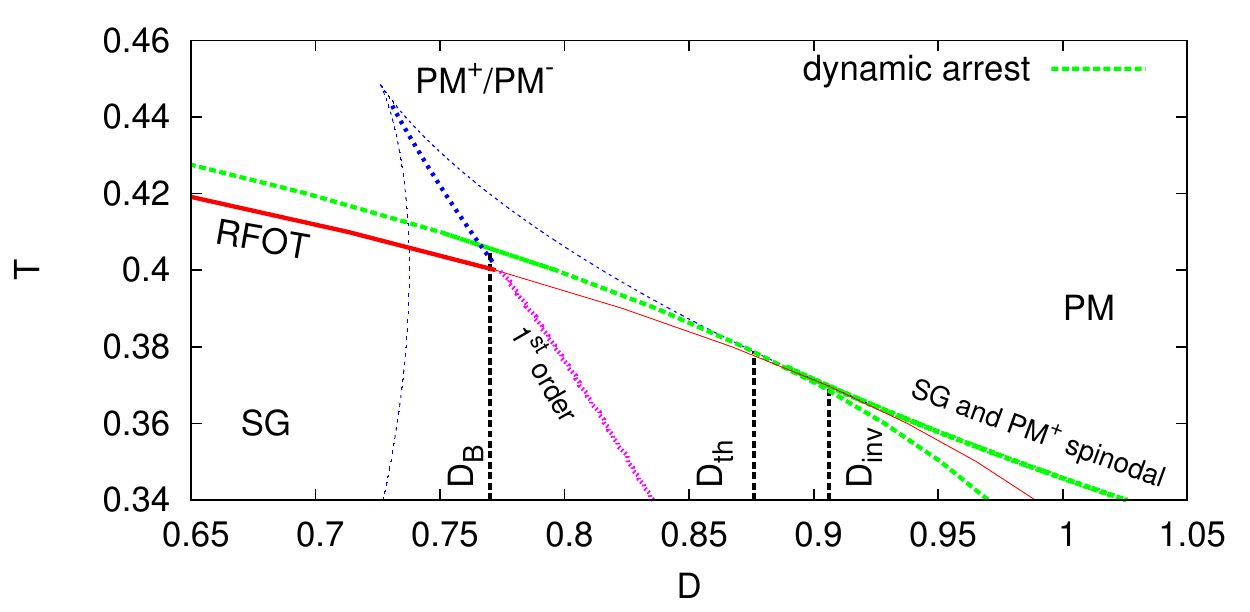}
\caption{The complete phase diagram for the model \refeq{H} for $p=3$. In the bottom inset the detail contained in the box of the main panel is displayed, together with he values of $D_B$, $D_{th}$ e $D_{\rm inv}$ (see text) identified by vertical dotted lines.}
\label{staticphasediagram}
\end{center}
\end{figure}

The resulting phase diagram for the model is reported in figure
\ref{staticphasediagram}.

From it, it can be seen that the system exhibits a rich phenomenology,
with both random first-order transitions (RFOT) and thermodynamic
first-order phase transitions (TFOPT). Here we will just comment on
them briefly to adequately introduce our dynamical study, along with
statics novelties with respect to previous analysis. The
interested reader can find all the details in \cite{Ferrari11}.

\paragraph{Random first-order transition}
For low enough $D$, the system exhibits the random first-order
phenomenology typical of the $p$-spin model. Along the dynamic
transition line $T_d(D)$, the system undergoes a \emph{dynamical
  arrest}, meaning that the relaxation towards the paramagnetic,
stable state is blocked by the presence of an exponentially large
number of metastable SG states which trap the dynamics. Being the SG
states metastable, this transition is not captured by the static
saddle-point equations, and has to be studied by solving the dynamics
of the model, or by using the marginality condition for the dynamics
\cite{Franz95,Crisanti08,Ferrari12,Crisanti92}.

At a temperature $T_s(D)$ \emph{lower} than $T_d(D)$, a \emph{static}
transition takes place, whereupon the number of SG states becomes
subexponential (their complexity, also called configurational entropy,
vanishes) and they become stable with respect to the paramagnet,
yielding the equilibrium SG solution \cite{Castellani05,Crisanti92}.
This feature of the model, occurring for $D<D_B=0.77$, is equivalent to
the $p$-spin model phenomenology. 

\paragraph{Thermodynamic first-order phase transitions}
In the region of the phase diagram
between the spinodal lines, but above the RFOT line, two paramagnetic
phases, termed PM$^-$ and PM$^+$, coexist, both with $q_1=0$ but with 
two different density values $d$: $d^-$ and $d^+>d^-$.
The two paramagnets are labelled $+$ and $-$ according to their
density value $d$ being, respectively, large and small. These values can
be determined by solving the saddle-point Eq. \refeq{eqsella2} in the
$q_1=0$ limit, yielding the expression
\begin{equation}
\frac{p}{2}d^{p}(1-d) = T^{2}(2d-1) + 2(TD-T^{2}\log2)(1-d)d.
\label{eqstatic}
\end{equation}

For $p=3$, this is a polynomial equation with 
three solutions for $d\in [0:1]$.
The solution with the intermediate value of $d$ turns out to be
 always unstable (see Ref. [\onlinecite{Ferrari11}]), leaving 
only a high-density and a low-density solutions.
 Since the
density $d$ has a continuous behavior along the PM$-$/SG thermodynamic
transition, Eq. \refeq{eqstatic} can be used, as well, to determine
the value of the density for the SG phase at the transition
point. This means that the spinodal lines $T(D)$ can be determined in a
parametric form in $d$ as 
\begin{eqnn}
D_{sp}(d) &=&\sqrt{\frac{p}{p-1}} \frac{d^{\frac{p}{2}-1}}
{2 (d-1) \sqrt{4 d^2-4 d+2}}\times
\\
\nonumber
&&\hspace*{-.5cm}
 \Bigl\{d^3 (p-1) 2\log (2)+d^2 \left[p (2-4 \log (2))+4\log (2)\right]
 \\
 &&\hspace*{-.4cm}+d \left[p (\log (4)-3)+1-2\log (2)\right]+p\Bigl\}
\nonumber
\\
\nonumber
&&
\\
T_{sp}(d) &=& \frac{(d-1) \sqrt{p(p-1)}d^{p/2}}{\sqrt{4 d^2-4 d+2}}.
\end{eqnn}
By studying those expressions, it can be readily checked that for
$p\geq3$, at high $D$ the spinodal curve is an asymptote of the $D$
axis; this means that, however large $D$, the system can always present
an high density phase, if the temperature is low enough. This fact
will be of capital importance in the following of this paper.

Besides the RFOT, thus, the system also exhibits both a PM$-$/PM$+$ and, furthermore, 
a PM$-$/SG \emph{thermodynamic} first order transition, that means
standard first order transitions with phase coexistence and latent
heat. Both transitions take place along the TFOPT line in figure
\ref{staticphasediagram} and coexisting phases exist
between the spinodal lines.
As the temperature is decreased to cross the RFOT dynamic
line, we observe that only the high-density
paramagnet undergoes dynamical arrest, while the low density PM$^-$ phase
is unperturbed. Thus, at $T$ lower than the crossing point of TFOPT and RFOT lines,
the transition occurs between a low density paramagnet PM$-$, with $d=d_-$
and $q_1=0$, and an high-density SG with $d=d_+$ and $q_1\neq0$. The
intersection takes place for $(D_B,T_B) = (0.77,0.40)$.

\paragraph{High density dynamical transition.}
At $(D_{\rm th},T_{\rm th})\simeq(0.876,0.379)$, the dynamical RFOT
line and the spinodal SG line intersect. It can then be seen that for
$D>D_{\rm th}$, the dynamical RFOT line coincides with the spinodal
line of the TFOPT, which means that the dynamical arrest in the PM$^+$
phase will take place as soon as phase separation occurs. From the
thermodynamic point of view, we have coexistence between two
paramagnets, as before. However, if we perform a quenching dynamics from the
high density PM phase, a dynamical arrest into a metastable SG
phase will take place.

\section{The dynamics \label{sec:dynamics}}
We are now ready to derive the dynamical equations for the model. 
Let us first separate the disordered part of the Hamiltonian
\refeq{H} from the deterministic one
\begin{eqn}
H & = & H_0  + H_J, \\
H_0 & = & (D-T\log2) \sum_{i=1}^{N}\frac{\s_{i}\tau_{i}+1}{2},\\
H_J & = & -\frac{1}{2^{p}}\pmexp{\sum}\pexp{J}(\s_{i{1}}+\tau_{i{1}})\dots(\s_{i{p}}+\tau_{i{p}}).
\end{eqn}
The relaxation dynamics is, then, governed by the $2N$ Langevin
equations
\begin{equation}
\begin{split}
\dot{\s}_i &= -\mu(t)\s_i(t) -\dpart{H_0}{\s_i}-\dpart{H_J}{\s_i}+ \eta_{i}(t),\\
\dot{\tau}_i &= -\nu(t)\tau_i(t) -\dpart{H_0}{\tau_i}-\dpart{H_J}{\tau_i}+ \theta_{i}(t),
\end{split}
\label{langevin}
\end{equation}
where we assume the noise fields $\eta_{i}(t)$ and $\theta_{i}(t)$ to be
delta-correlated:
\begin{eqn}
\left<\eta_{i}(t)\eta_{j}(t')\right> = \left<\theta_{i}(t)\theta_{j}(t')\right> &=& \delta_{ij}D_{0}(t-t'),\\
\left<\eta_{i}(t)\theta_{j}(t')\right> &=& 0,
\end{eqn}
with (taking the Boltzmann constant $k_{B}=1$)
$$
D_{0}(t-t') = 2T\d(t-t').
$$ 
Following \cite{Crisanti93}, we have inserted the Lagrange
multipliers $\mu(t)\s_i(t)$ and $\nu(t)\tau_i(t)$ in order to enforce
the spherical constraint.

The quantities we are interested in are the correlation functions and
the response functions of the system. In this case, differently from
the standard spherical p-spin model, we have two different types of
degrees of freedom, and thus two possible external perturbing fields,
one for each of them. As a result of this, we have four different
correlation functions
\begin{eqn}
C_{\s\s}(t,t') = \frac{1}{N}\sum_{i=1}^{N}\dav{\thav{\s_{i}(t)\s_{i}(t')}}, \\
C_{\s\tau}(t,t') = \frac{1}{N}\sum_{i=1}^{N}\dav{\thav{\s_{i}(t)\tau_{i}(t')}},\\
C_{\tau\s}(t,t') = \frac{1}{N}\sum_{i=1}^{N}\dav{\thav{\tau_{i}(t)\s_{i}(t')}},\\
C_{\tau\tau}(t,t') = \frac{1}{N}\sum_{i=1}^{N}\dav{\thav{\tau_{i}(t)\tau_{i}(t')}},
\end{eqn}
and four response functions
\begin{eqn}
R_{\s\s}(t,t') = \frac{1}{N}\sum_{i=1}^{N}\dfunc{\dav{\thav{\s_{i}(t)}}}{h_{i}(t')}, \\
R_{\s\tau}(t,t') = \frac{1}{N}\sum_{i=1}^{N}\dfunc{\dav{\thav{\tau_{i}(t)}}}{h_{i}(t')},\\
R_{\tau\s}(t,t') = \frac{1}{N}\sum_{i=1}^{N}\dfunc{\dav{\thav{\s_{i}(t)}}}{l_{i}(t')},\\
R_{\tau\tau}(t,t') = \frac{1}{N}\sum_{i=1}^{N}\dfunc{\dav{\thav{\tau_{i}(t)}}}{l_{i}(t')},
\end{eqn}
where $h_{i}(t)$ and $l_{i}(t)$ are time-dependent perturbing fields
conjugated with the $\s_{i}$ and $\tau_{i}$ degrees of freedom,
respectively.

Our aim is to use equations \refeq{langevin} to obtain self-consistent
dynamical equations for the functions above. We
employ the generating functional method devised in \cite{Martin73} by
Martin, Siggia and Rose, and already used for the $p$-spin model in
\cite{Kirkpatrick87} by Kirkpatrick and Thirumalai. The MSR approach
consist essentially in defining a generating functional
$Z[\bh,\bl,\bhh,\bhl,J_{ij}]$ \footnote{In order to lighten the
  notation, in the following we shall use boldface to denote lattice
  site-dependent quantities. For example $\bs = (\s_{i})_{i=1}^{N}$,
  and $\bs\cdot\bs =\sum_{i}\s_{i}\s_{i}$} for the $2N$
$1$-dimensional random fields $\s_{i}(t)$ and
$\tau_{i}(t)$\footnote{Good introductions to the MSR formalism can be
  found in Refs. \cite{Castellani05,DeDominicis06}.}. The correlation
and response functions can, then, be obtained by taking functional
derivatives of $Z$ with respect to the external fields
$\bh(t),\bl(t),\bhh(t)$ and $\bhl(t)$, as in an usual field theory.

We have emphasized the fact that the generating functional still
depends on the quenched random couplings $J_{ij}$, and so does every
quantity generated by it; so, in principle, we would have to average
them over the disorder in order to obtain the correlation and response
functions we want. However, as remarked by De Dominicis in
\cite{DeDominicis78}, since the generating functional in absence of
external currents is by definition normalized to one
$$
Z[0,0,0,0,J_{ij}]=1,
$$
it is independent from the variables of the system, and so it can be
averaged over the disorder directly. This is in contrast with the
static partition function for a system with quenched disorder, which
is not self-averaging. As a result of this, the use of replica theory
is not needed in the dynamical framework; this fact constitutes the
main advantage of the dynamical approach over the static one.

Performing the average over the disorder leads to a decoupling of the
lattice sites and a coupling of the configurations of the system at
different times, as it happens for the $p$-spin model in
\cite{Kirkpatrick87}. This is to be conceptually compared to the
decoupling of the sites in the static approach with replicas, yielding
a coupling between different replicas \cite{Gross84}.  It is then
possible, using saddle point methods, to write an effective generating
functional which yields two single site dynamic equations valid for every
degree of freedom in the lattice.  The details of the derivation of
the dynamics can be found in the appendix. We report here the
site-independent dynamical equations
\begin{subequations}
\begin{eqnn}
\dot{\s} &=& -\mu\s(t)-\frac{(D-T\log2)}{2}\tau(t) \label{myeqsigma} \\ 
&&+K_{p}(p-1)\int dt'\ \R(t,t')\C(t,t')^{p-2}(\s(t')+\tau(t'))\nonumber\\
&& + \xi(t),\nonumber
\end{eqnn}
\begin{eqnn}
\dot{\tau_i} &=& -\nu\tau(t)-\frac{(D-T\log2)}{2}\s(t)\label{myeqtau} \\ 
&&+K_{p}(p-1)\int dt'\ \R(t,t')\C^{p-2}(t,t')(\s(t')+\tau(t'))\nonumber\\ 
&&+ \zeta(t),\nonumber
\end{eqnn}
\end{subequations}
where the correlation matrix for the noise terms $\xi(t)$ and
$\zeta(t)$ has been renormalized in the following way
\begin{subequations}
\begin{equation}
\begin{split}
\left<\xi_{i}(t)\xi_{j}(t')\right> =& \d_{ij}K_{p}\C(t,t')^{p-1} + \d_{ij}2T\delta(t-t')),\\
=&\left<\zeta_{i}(t)\zeta_{j}(t')\right>\\
\left<\xi_{i}(t)\zeta_{j}(t')\right> =& \d_{ij}K_{p}\C(t,t')^{p-1}. \label{corr}
\end{split}
\end{equation}
\end{subequations}
We have also defined the constant
$$
K_{p} \equiv \frac{J^{2}p}{2^{2p+1}}
$$
and the two functions
\begin{subequations}
\begin{eqnn}
\C &\equiv& C_{\s\s}+C_{\s\tau}+C_{\tau\s}+C_{\tau\tau},\\
\R &\equiv& R_{\s\s}+R_{\s\tau}+R_{\tau\s}+R_{\tau\tau}.
\end{eqnn}
\end{subequations}

The equations for the correlation and response function can then be
easily obtained, as reported in appendix \ref{appendA}.

\subsection{Symmetries, equilibrium and ergodicity}
In appendix \ref{appendA}, we derive eight coupled differential
equations for eight different unknown functions. We now specify them to 
the particular problem we want to study, i.e. identifying dynamical arrest.
In order to to this, we can restrict ourselves
to an equilibrium (\emph{i.e.} starting from an equilibrium initial condition) and ergodic dynamics. This implies time-translational invariance (TTI) of the correlators 
\begin{equation}
C(t,t') = C(t-t')
\end{equation}
and that the fluctuation-dissipation theorem (FDT) holds
\begin{equation}
R(t) = - \theta(t)\frac{1}{T}\frac{dC(t)}{dt},
\end{equation}
where $R$ and $C$ denote any correlation and response function couple in the
system, respectively, and $\theta(t)$ is the Heaviside step
function. These assumptions are valid in the high temperature PM
phase, where ergodicity is not broken, but they are generally false
when the system is cooled below the dynamical transition temperature
$T_{d}$, where a transition to a $SG$ phase with broken ergodicity
takes place. Second, we notice that both the model Hamiltonian
\refeq{H} and the effective generating functional \refeq{myendgen} are
symmetric with respect to a $\bs\leftrightarrow\bt$ switch
$$
H(\bs\rightarrow\bt,\bt\rightarrow\bs) = H(\bs,\bt).
$$ 
This means that the $\bs$ and $\bt$ evolve in the same statistical
ensemble, which, in turn, implies that the correlation functions obey
the relations
\begin{equation}
C_{\s\s}(t,t') = C_{\tau\tau}(t,t'), \quad C_{\s\tau}(t,t') = C_{\tau\s}(t,t'),
\end{equation}
that can then be extended to the response functions by exploiting the FDT
\begin{eqnn}
R_{\s\s} = -\theta(t)\frac{1}{T}\frac{dC_{\s\s}}{dt} &=& -\theta(t)\frac{1}{T}\frac{dC_{\tau\tau}}{dt} = R_{\tau\tau},\\
R_{\s\tau} = -\theta(t)\frac{1}{T}\frac{dC_{\s\tau}}{dt} &=& -\theta(t)\frac{1}{T}\frac{dC_{\tau\s}}{dt} = R_{\tau\s}.
\end{eqnn}

Once that these relations are established, we can see that only the two
correlation functions $C_{\s\s}$ and $C_{\s\tau}$ are needed to
completely describe the dynamics of the system.  Thus, we can define
the ``total'' correlation function of the system $C(t-t')$, by
normalizing the $\C$ to one
$$
C(t-t') = \frac{\C(t-t')}{\C(0)}=\frac{2(C_{\s\s}(t-t')+C_{\s\tau}(t-t'))}{4d(0)}.
$$
where we have used the spherical constraint
$$
\sum_{i=1}^{N}\thav{\s_i(t)\s_i(t)} = N  \Longrightarrow C_{\s\s}(t,t) = 1
$$
and the relation \refeq{trasf} between the $\s\tau$ product and the site occupation number
\begin{eqnarray}
\frac{\s_{i}(t)\tau_{i}(t)+1}{2} &=& n_{i}(t)
\\
\Longrightarrow C_{\s\tau}(t,t) = C_{\tau\s}(t,t) &= &2d(t)-1.
\end{eqnarray}
Since we assume the dynamics to be at equilibrium, the density $d(t)$,
being an one-time quantity, is a constant of motion, equal to its
equilibrium value
\begin{equation}
d(0) = d(t)= d(\infty) \quad\forall t.
\label{densitycondition}
\end{equation}

Using the equations for the $C_{\s\s}$ and $C_{\s\tau}$, we obtain the
following integro-differential equation for the total correlation
function $C(t)$
\begin{eqnn}
\dot{C}(t) &=& \mathcal{A}(D,T,d) C(t) \label{eqdinamica}
\\
&& - \mathcal{B}(T,d) \int_{0}^{t}du\ C^{p-1}(t-u)\dot{C}(u) \nonumber
\\
\mathcal{A}(D,T,d)&\equiv& \left(D-T\log2 -\frac{J^{2}d^{p-1}p}{4T}\right)(d-1) -T 
\nonumber
\\
\mathcal{B}(T,d)&\equiv &\frac{J^{2}p d^{p-1}}{4T}
\nonumber
\end{eqnn}
which is a closed equation, whose solution can be found once the
values of the parameters $D$, $T$ and $d$ are known. 
It can be seen as a Mode Coupling schematic equation, and it is worth noticing that in the  limit 
of the  $d$ parameter (the fraction of filled-in sites) going to $1$
one recovers standard MC equations.
The density $d$ is
assumed, at all times, to be equal to its equilibrium value, given by
Eq. \refeq{eqstatic}.
As a result of this, in the region between spinodal lines, where phase
coexistence occurs, the dynamics will have to be studied separately
for each one of the two coexisting phases. 
In the high density limit case $d\to 1$, thermodynamically occurring for low $D$, the proper limit 
of Eq. \refeq{eqdinamica} is considered in appendix  \ref{sec:dto1}.
The study of Eq. \refeq{eqdinamica} will be one of the main focuses of this work.

An equation can be derived also for the difference $\Delta C$ between $C_{\s\s}$
and $C_{\s\tau}$, obtaining
\begin{equation}
\frac{d(\Delta C(t))}{dt}= \left[(D-T\log 2)d-\frac{p}{4T}d^p -T\right]\Delta C(t),
\label{A}
\end{equation}
which has the trivial solution
\begin{equation}
\Delta C(t) = 2(1-d)e^{At}
\label{eqdinamicadiff}
\end{equation}
where the $A$ constant is defined as
$$
\qquad A \equiv (D-T\log 2)d-\frac{p}{4T}d^p -T.
$$
If now we use Eq.  \refeq{eqstatic} to eliminate $D$, we get
$$
A = -\frac{T}{2(1-d)}
$$
which is always negative for any value of $d$ and
$T$. This means that the difference between the two correlators tends
to zero for long enough times, and thus the dynamics of the system can
be always solved using the $C(t)$ function only.

\subsection{Solving the dynamic equations}

In this section we report the results obtained by numerically solving 
equation \refeq{eqdinamica} in various representative points of the
phase diagram. Eq. \refeq{eqdinamica} is an integro-differential
mode-coupling like equation~\cite{Goetze09}, that can be solved using
the standard algorithm intruced by Fuchs \etal \cite{Fuchs91} and extended in different ways, cf. e. g., Refs.
\cite{Berthier07,Crisanti11,Crisanti15}. We consider three meaningful cases to illustrate the varoius occuring regimes.

\subsubsection{$D=0$: dynamic transition}
\begin{figure}[t!]
\begin{center}
\includegraphics[width = 0.7\columnwidth,angle=270]{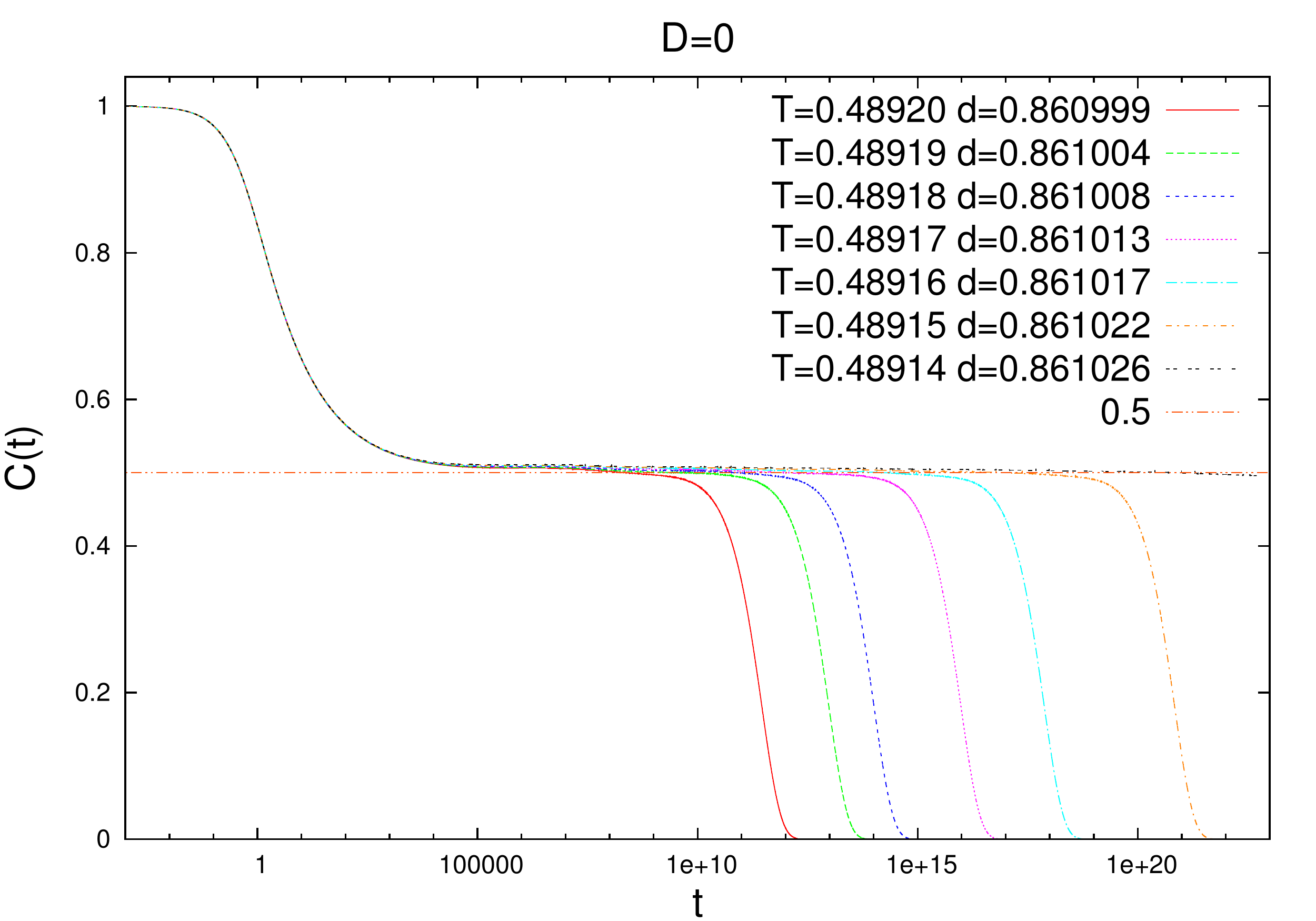}
\caption{The total correlation function $C(t)$ evaluated for various
  temperatures in the vicinity of the dynamical transition at
  $T_d(D=0)=0.4892$; for each temperature, the value of the
  equilibrium density $d$ of the system is reported. As the system is
  cooled, the function develops a plateau whose length increases
  rapidly with decreasing $T$, to become infinite at the dynamical
  transition line.}
\label{D=0}
\end{center}
\end{figure}

At $D=0$ and high $T$ the system yields a single paramagnetic phase
(PM), which undergoes a dynamical transition with ergodicity breaking at
$T_d(D=0)=0.4892$ \cite{Ferrari11}. In figure \ref{D=0} we
report the total correlation function of the system as the transition
is approached from above.

As we can see, our dynamic equation \refeq{eqdinamica} yields the
dynamic transition predicted in \cite{Ferrari11}, and the correlator
shows the typical phenomenology of a mode-coupling like dynamic
arrest. The transition temperature $T_d$ corresponds up to a
$O(10^{-4})$ error with the predicted value $0.4892$ and the height of
the plateau $q_d$ (also called the non-ergodicity parameter) is
 $(p-2)/(p-1)=0.5$, as expected.

\subsubsection{$D=D_B$: phase coexistence}
\begin{figure}[t!]
\begin{center}
\includegraphics[width = 0.7\columnwidth,angle=270]{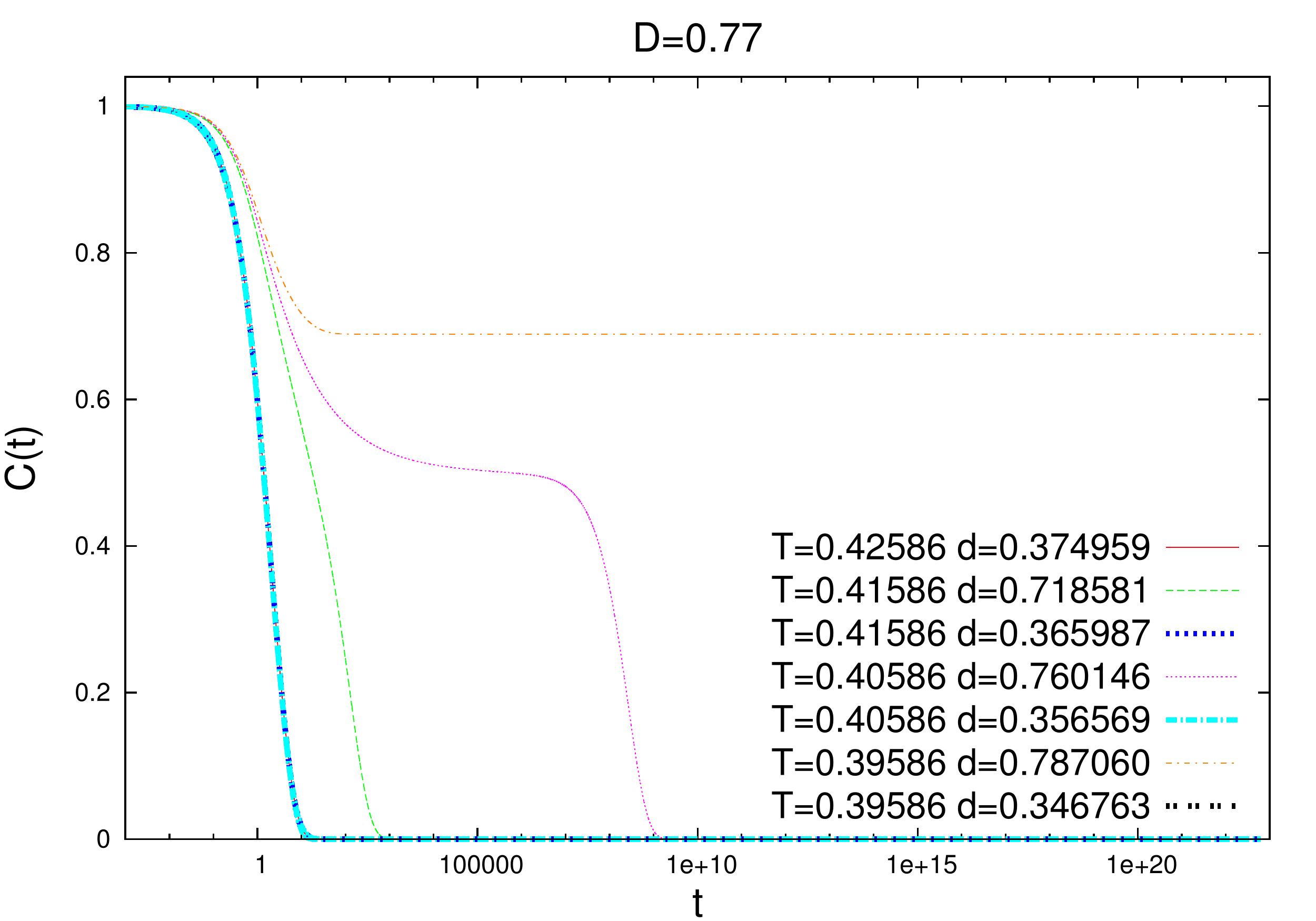}
\caption{The total correlation function $C(t)$ evaluated for various temperatures as the system enters the phase coexistence zone and the high density phase undergoes dynamical arrest. Here $T_d(D=0.77) \simeq 0.40586$. In the phase coexistence zone, the PM$+$ correlator is plotted with a thin line, while the PM$-$ with a thick one. As we can see, the PM$+$ makes a transition to an arrested SG phase while the PM$-$ remains paramagnetic.}
\label{D=0.77}
\end{center}
\end{figure}

For $D=D_B=0.77$, the situation is richer and more interesting; for
this value of the crystal field, at low  temperature the
system undergoes phase coexistence, yielding two separate paramagnetic
phases with high (PM$^+$) and low (PM$^-$) density. As we anticipated,
the dynamics of the system has to be solved separately for each one
of these two phases. Their behavior turns out, actually, to be 
quite different, as only the high density PM$+$ phase 
undergoes ergodicity breaking as the dynamical line is crossed. In
figure \ref{D=0.77} we plot the resulting correlators as the system
enters the phase coexistence zone, and the high density phase
undergoes the dynamic transition.

We see that, again, the expected phenomenology is reproduced by our
equation. At $T=0.42586$ only a single paramagnetic phase is present,
but for $T=0.41586$, two different paramagnetic phases separate; at
$T=0.40586\simeq T_d(D=0.77)$, the high density phase correlator shows
the typical plateau as the dynamic transition is approached, and as
the cooling continues, the high density PM$^+$ phase eventually undergoes a
dynamical arrest, while the low density paramagnet PM$^-$ stays
unchanged until the first order phase transition takes place to the thermodynamic 1RSB-stable glass phase.

\subsubsection{$D>D_{th}$: anomalous dynamical arrest}
Up to the prevuoius case, our dynamic analysis confirms the results of the static one
\cite{Ferrari11}. However, as we anticipated in section
\ref{sec:intro}, for $D>D_{th}$, our dynamical equation yield dynamic
arrest at finite temperature when starting from the PM$^+$ phase,
an effect that was not identified  in the static analysis, where the thermodynamically dominant phase is low 
density PM$^-$. We
shall now report the results obtained by solving equation
\refeq{eqdinamica}, leaving the study and discussion of the new
arrested phase for the next section.

\begin{figure}[t!]
\begin{center}
\includegraphics[width = 0.7\columnwidth,angle=270]{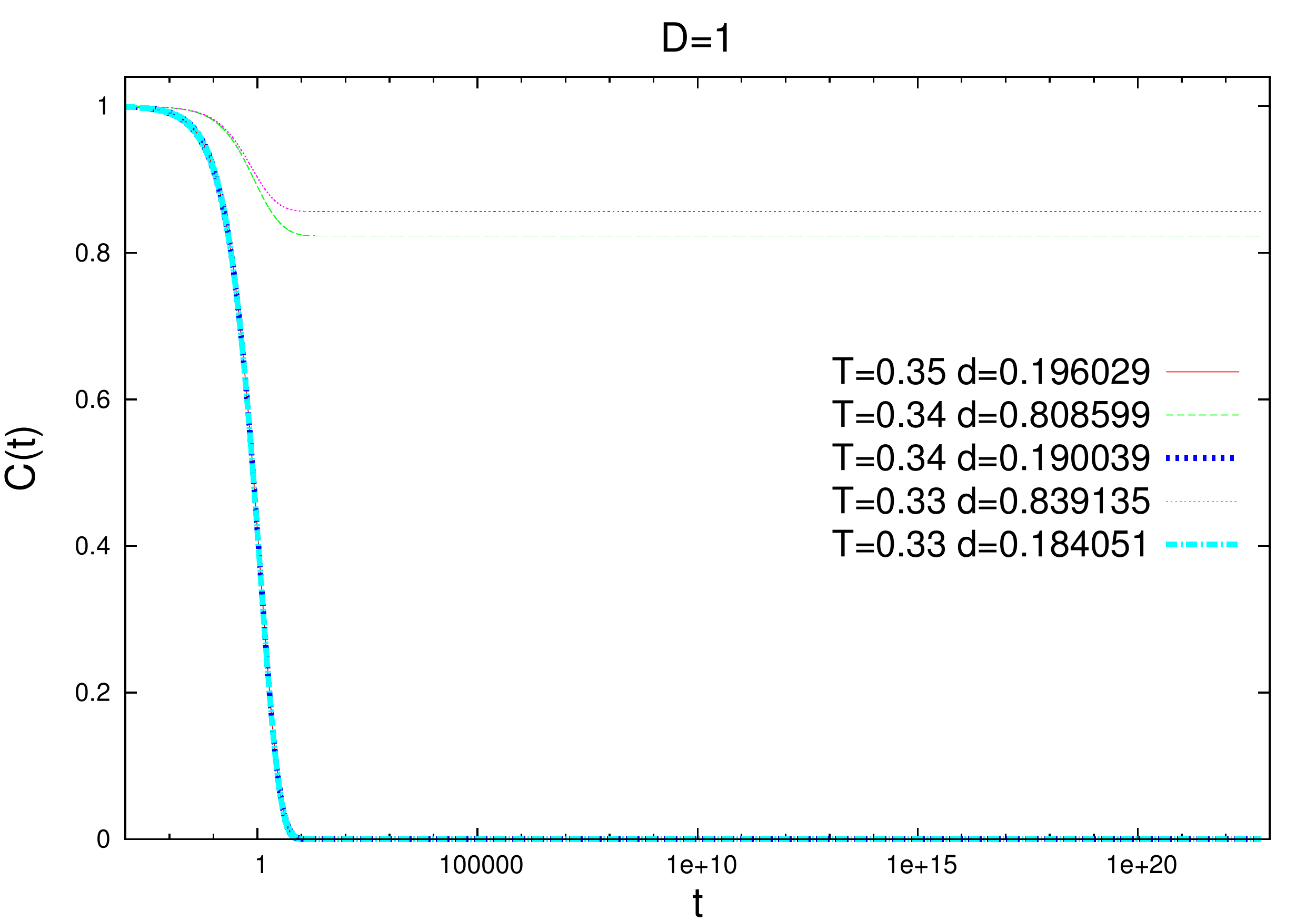}
\includegraphics[width = 0.7\columnwidth,angle=270]{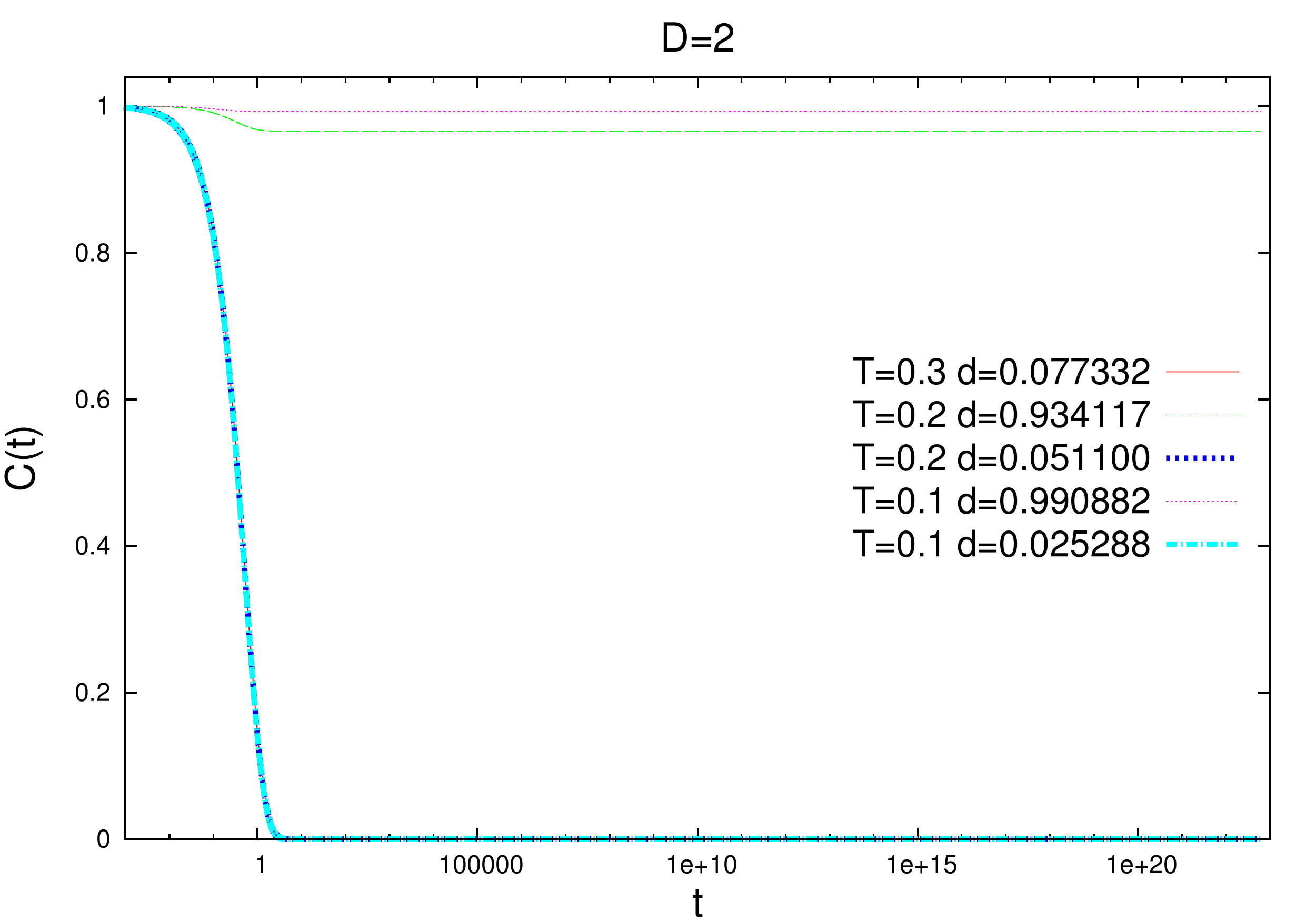}
\caption{The total correlation function $C(t)$ evaluated for various
  temperatures as the system enters the phase coexistence zone and the
  anomalous dynamic transition takes place. The presence of this
  dynamic transition is a novelty with respect to the results of
  \cite{Ferrari11}.}
\label{D=1}
\end{center}
\end{figure}

In order to show the onset of this new transition, we choose
$D=1>D_{th}$, $T=0.35$ and we cool the system starting from a high density initial condition
 crossing the spinodal
line  during the
procedure.  The PM$^+$ spinodal line, for these chemical potential values,  turns to be a dynamical arrest transition line, as shown  in figure \ref{D=1}. We can observe that, as
soon as the two PM phases separate, the high density phase is already
an arrested SG phase, with a nonzero overlap, while the low density phase shows no sign of dynamical arrest. 
This is at difference with
respect to the static results of Ref. \cite{Ferrari11}, where the high density phase is supposed to be still
paramagnetic for $T=0.34$.

What is puzzling about this result is the fact that the high density
phase is already deep into the SG when the separation occurs: for
$T=0.34$ we have $q_d\approx0.82$, which is already much higher than
the $(p-2)/(p-1)=0.5$ that we would expect for a system which approaches
from above an usual dynamical arrest. This means that if the high
density phase existed even above the spinodal line, its dynamic
transition temperature would be actually much higher than $T=0.34$;
however, this effect is not visible since only the low density PM$^-$
phase exists in that region. If we follow the same cooling procedure for
$D=2$ (i.e., we cross the spinodal line at $D=2$ during the cooling),
the results are not very different, and the high density arrested phase
is still present. In this case, the value of the overlap at the
separation line is even higher, with $q_d \approx 0.97$ for $T=0.2$.

Solving equation \refeq{eqdinamica} for higher $D$ does not change the
general situation, so we will not report any results for higher
values. The point is that, since the spinodal line is an asymptote of
the $D$ axis (as we have mentioned in section \ref{sec:statics}), then
for arbitrarily large $D$, a high density phase exists at
$T>0$. According to both the marginality condition for the statics,
and the dynamic results presented so far, this phase presents a
dynamical arrest into a SG phase with nonzero overlap, that can be realized by selecting atypically 
dense initial conditions at those temperature and chemical potential.

\section{Complexity and free energy}
\begin{figure}[htb!]
\begin{center}
\includegraphics[width = 0.75\columnwidth]{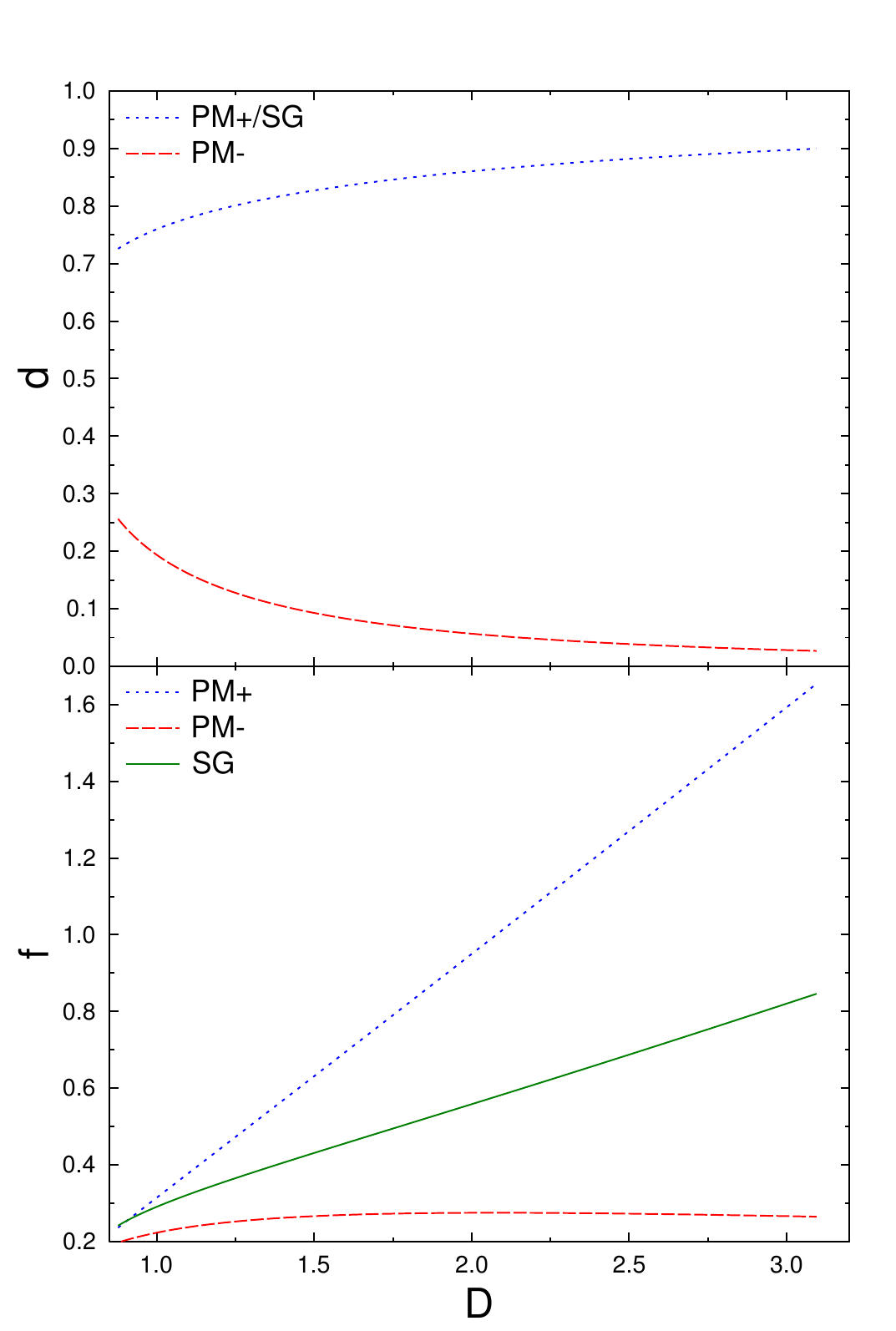}
\caption{The state free energy of the low density paramagnet (PM-),
  the high density paramagnet (PM+), and of the spin glass states
  which trap the dynamics (SG), evaluated along the high density dynamical
  arrest line for $D>D_{th}=0.876$. We also report the densities of the two
  paramagnets. We see that the free energy $f$ of the trapping states becomes equal to the one of the 
  paramagnet for $D=D_{inv} \simeq 0.9062$}
\label{fig:freeenergy}
\end{center}
\end{figure}

\begin{figure}[htb!]
\begin{center}
\includegraphics[width = 0.9\columnwidth]{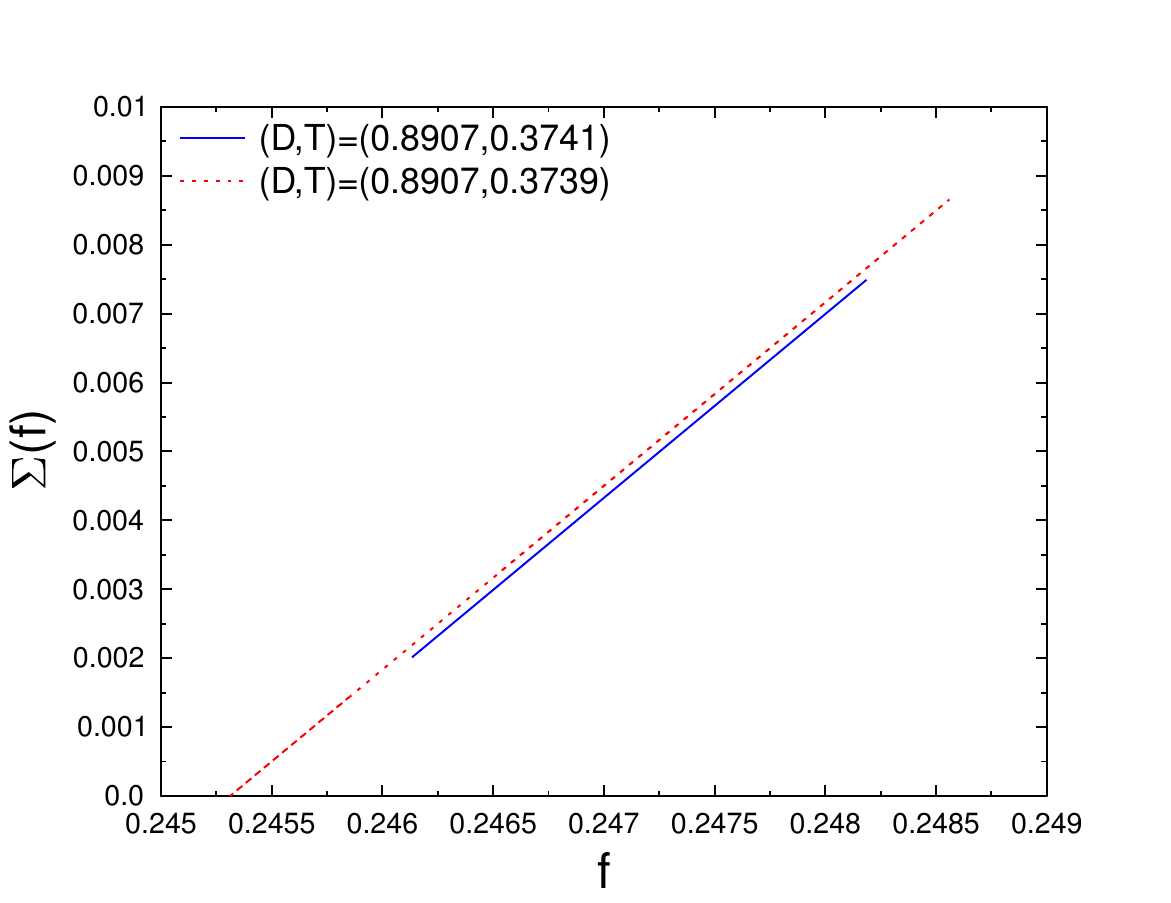}\\
\includegraphics[width = 0.9\columnwidth]{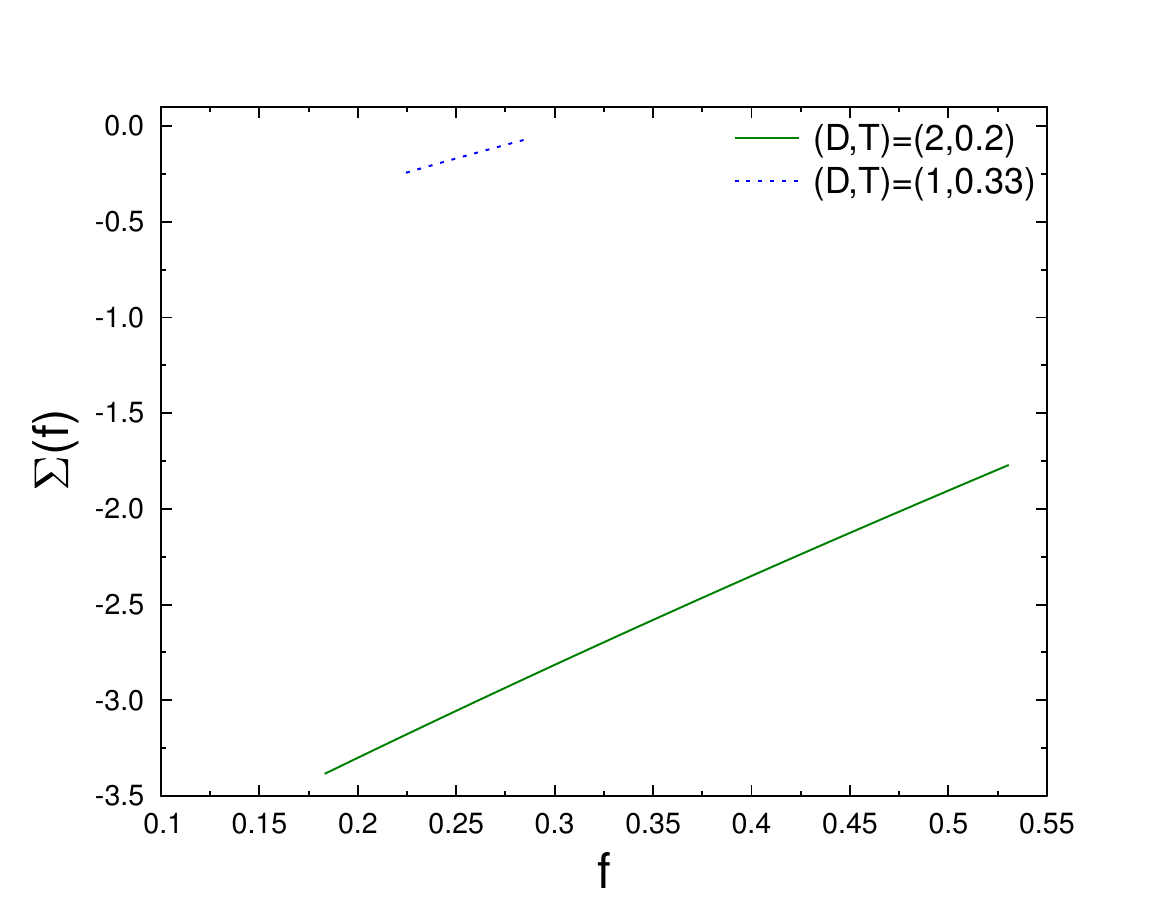}
\caption{The complexity of the states which trap the dynamics, below
  the high density dynamical arrest line. Upper Panel: complexity for
  two points with $D<D_{inv} \simeq 0.9062$, with
  $T_s(D)<T<T_d(D)=...$ (full line) and $T=T_s(D)=...$ (dotted line),
  respectively. Lower panel: complexity for two representative points
  with $D>D_{inv}$.}
\label{complexity}
\end{center}
\end{figure}

Since the system undergoes a dynamical arrest, we would expect the RFOT
phenomenology which holds in the other regions of the phase diagram to
be present in this case as well, inside the high $d$ minimum which
corresponds to the PM+ phase (the low density paramagnet is completely
orthogonal to our discussion). In summary, we expect the metastable
states which trap the dynamics (and maximize the complexity) to have
a higher in-state free-energy than the one of the paramagnetic,
ergodic state. We might also expect a complexity $\Sigma(f)$ to be
strictly positive for every $f$ up to a static temperature $T_s(D)$
where states with null complexity are born and a static transition
takes place. We computed both quantities using replica theory as
in Ref. \cite{Ferrari11} and report the corresponding curves in figures
\ref{fig:freeenergy} and \ref{complexity}.
We see that the usual RFOT picture does not hold for all values of $D$
along the new dynamical arrest line. Indeed, it holds only for
$D<D_{inv}$ where $D_{inv} \simeq 0.9062$, which in fact corresponds
to the point where the static line touches the new dynamical arrest
line. In this region, the complexity of the states is positive for
every $f$ (as one can see in the upper panel of figure
\ref{complexity}) and the free energy of the $m=1$ states which trap
the dynamics is higher than the paramagnetic one. On lowering $T$ at
constant $D$, the complexity touches zero (corresponding to the
condition Eq.  \refeq{eqsella3}) and a static transition to a SG phase
takes place as usual.

This picture does not hold anymore for $D \geq D_{inv} \simeq 0.9062$.  
For $D=D_{inv}$, the maximum complexity condition
(dynamical arrest line) and the zero complexity condition (static line)
coincide, which means that the complexity is zero (i.e., the number of
all states becomes subexponential) and the free energy of the trapping
states becomes equal to the paramagnetic (PM+), cf. figure
\ref{fig:freeenergy}.  This is alike to the occurrence of the static
Kauzmann transition but in this case both SG and PM$^+$ phases are
metastable and dynamically occurring only for initial conditions with
a density atypically high for these chemical potential $D$ values.  When
$D$ is increased along the high density arrest line, the free energy
of the trapping states becomes lower than the paramagnetic one and
continues to decrease indefinitely, while the complexity
becomes more and more negative as reported in figure
\ref{complexity}. If one looks at the phase diagram
\ref{staticphasediagram}, it is possible to see that there is a range
of $D$ such that, on cooling, the static line is met before the
dynamic one \cite{Ferrari11}.  This would mean that along this path in
the parameter space the states with low free energy and complexity
arise before the ones with maximum complexity and free energy, only
eventually appearing on the maximum complexity line (that, for
$D<D_{th}$, is the dynamic line): the complexity of the glassy
metastable states becomes again non-zero crossing the static line on cooling
and reaches its maximum at the line formerly denoted as dynamic
line. However, this is true for $D>D_{th}\simeq 0.876$ up until
$D\sim 1.6$. From that point on, the
static line touches the $T=0$ axis and the complexity stays negative
for every value of $D,T$. In summary, using both the dynamical equations and 
the replica approach, we find an anomalous, complexity free,
dynamical transition,  that occurs if the system is initially prepared at values of the density
corresponding to a paramagnetic metastable state at high density. This state always
 coexists with a more probable and
thermodynamically dominant low density paramagnetic state.
It is the latter which, at lower $T$ and/or lower $D$,
undergoes a first order phase transition to a spin-glass phase.

\section{Conclusions}

In this work, we have studied, both dynamically and statically, a disordered model which
shows both an RFOT-like phenomenology (dynamical arrest, complexity, Kauzmann transition, etc.), and 
an ``ordinary" first order phase transition with latent heat and phase coexistence. We have derived and solved the equations
for the equilibrium dynamics of the model and completed the static replica-based study of \cite{Ferrari11} with novel results.
In doing so, we have noticed the presence, in a certain region of the phase diagram, of a RFOT-like dynamical arrest line,
which however shows a non-positive complexity and does not work as a precursor for a Kauzmann-like static transition, in contrast with the usual
phenomenology expected in RFOT models.

The picture we propose to explain this is the following. Our model has
two order parameters, the density $d$ and the 1RSB self-overlap $q_1$
(termed $q$ from now on). One can imagine to construct a potential
function of $d$ and $q$, by plotting the paramagnetic free energy as a
function of $d$, and then performing a Franz-Parisi like
\cite{Franz95} construction along the $q$ axis, for every $d$.  For
$D<D_B=0.77$ the transition scenario is qualitatively alike to the random
first order transition one. 

For chemical potential  values in the interval 
 $D_B<D<D_{th}$, instead, the scenario changes. For high $T$ only one minimum is present but, on
cooling, the spinodal line of a second paramagnetic phase, at {\em
  higher} density, is crossed:
 a second, metastable minimum with $q=0$ and density $d=d^+$ is
formed, termed $PM^+$.  Further lowering  the temperature the dynamical transition line is crossed. 
There, a metastable phase with $q\neq0$ and $d=d^+$ arises from the PM$^+$ solution, 
with a higher free energy. 
This phase consists of many equivalent states
and corresponds to the arrested glassy phase. The difference in free energy
between the SG metastable minima and the free energy of the PM$^+$ minimum
 is equal to the complexity counting the log of the number of  SG metastable states. If $T$ is
lowered further, we cross the static transition line, where the $q\neq
0$ minimum has the same height as the paramagnetic one (null
complexity) and thus it becomes stable: we have a static transition in
a spin-glass phase. 

To summarize, we have a three-step process on cooling for $D\in [D_B,D_{\rm th}]$:
\begin{enumerate}
 \item A secondary PM$^+$ minimum with $d=d+$ and $q=0$ is formed (the stable phase is the low density PM$^-$).
 \item A SG minimum with $d=d+$, $q\neq 0$ is formed, arising from PM$^+$.
 \item The SG becomes stable with respect to the low density 
paramagnet PM$^-$ and a static transition takes place.
\end{enumerate}
This scenario almost corresponds the usual RFOT phenomenology. The only difference is that there are two paramagnetic phases, one stable (PM$^-$) and one metastable at higher density, PM$^+$, and that the glassy metastable states
at the threshold free energy arise \emph{inside the
PM$^+$}.

When $D$ is raised beyond $D_{\rm th}$, the order of these steps above is
changed. Two apart scenarios appear.
For $D_{\rm th}<D<D_{\rm inv}$, steps $1$ and $2$ exchange. Fixing
 $T$ slightly above $T_{\rm sp}(D)$, $d=d^+$ and looking at
the potential only in the direction of $q$, the $q\neq 0$ minimum will
have already formed. However, this has no effect on the
thermodynamics since $d=d^+$ does not yet correspond to a minimum on
the $d$ axis. For that to happen, $T$ must be lowered to cross the PM$^+$ spinodal.
At lower $T$ the process goes on as previously.

For $D>D_{\rm inv}$, step 1 becomes the last to happen: again, the system dynamics is arrested as soon as the 
PM$^+$ spinodal is reached, but the SG minimum along $q$
not only is formed before the PM$^+$ has a chance to appear, but it has
even become \emph{stable} with respect to the  PM$^+$. 
We stress the fact that this unusual
behavior is possible only because the system has two order
parameters, differently from usual RFOT models whose behavior is
governed only by $q$. The fact that the minimum in $q$ is already
formed for $D>D_{\rm th}$ is, indeed, confirmed by the fact that the naive
marginality condition $d^2F/dq^2=0$,  used in Ref.
\cite{Ferrari11},
completely  misses this new arrest line: in the direction of $q$ the minimum is
already formed and the curvature is positive. The whole Hessian matrix
of the replicated free energy with respect to $q$ and $d$ (evaluated
at $m=1$) must be used to detect the new line, as reported in Sec. \ref{sec:newstatics}.
This picture is corroborated by the fact that this
phenomenology is found in the PM$^+$ phase, whose density goes up and
approaches $1$ as $D$ increases, as reported in the upper panel of
figure \ref{fig:freeenergy}: this means that the PM$^+$ phase of our
model becomes more and more similar to the usual $p$-spin spherical
model (PSM) \cite{Crisanti92}, and the dynamical equations behave
accordingly as explained in appendix \ref{sec:dto1}. 
However, we also stress that such density values are
 thermodynamically extremely unlikely to occur for these large
values of the chemical potential $D$. 



\section{Acknowledgements}
The research leading to these results has
received funding from the Italian Ministry of Education, University
and Research under the Basic Research Investigation Fund (FIRB/2008)
program/CINECA grant code RBFR08M3P4 and under the PRIN2010 program,
grant code 2010HXAW77-008, and from the European Research Council, under the European Union's Seventh Framework Programme
FP7/2007-2013/ from the People Programme (Marie Curie
Actions) under REA grant agreement n 290038, NETADIS project, and ERC grant agreement n 247328, CryPheRaSy project.
U.F. thanks A. Destexhe for hosting at the European Institute for Theoretical Neuroscience.
\clearpage

\appendix

\section{Derivation of the dynamical equations for the p-spin Blume-Capel spherical model\label{appendA}}
In this appendix we provide the step-by-step derivation of the dynamical equations studied in this paper.

\subsection{Computation of the effective generating functional}

We write down the full expression for the MSR generating functional
\begin{widetext}
\begin{equation}
\begin{split}
Z[\bh,\bl,\bhh,\bhl,J_{ij}] = &\int \D\s\D\hat{\s}\D\tau\D\hat{\tau}
\times\exp\Bigg[ -\frac{1}{2}\int dt dt'\bsh(t)\cdot\bsh(t)D_0(t-t') +\bth(t)\cdot\bth(t')D_0(t-t')\Bigg]\\
\times &\exp\left[\int dt\ i\left(\bsh(t)\cdot(\bm{\dot{\s}}(t) + \mu(t)\bm\s(t) + \bm{\partial_{\s}H_{0}}(t)) + \bm{\th}\cdot(\bm{\dot{\tau}}(t)+ \nu(t)\bm{\tau}(t) + \bm{\partial_{\tau}H_{0}}(t))\right) \right]\\ 
\times &\exp \left[\int dt\ i\left(\bsh(t)\cdot\bm{\partial_{\s}H_{J}}(t)) + \bth(t)\cdot\bm{\partial_{\tau}H_{J}}(t)\right) \right],\\
\times &\exp\left[-\int dt\ i(\bh(t)\cdot\bsh(t)+\bhh(t)\cdot\bs(t)+\bl(t)\cdot\bth(t)+\bhl(t)\cdot\bt(t))\ \right]
\end{split}
\label{Z}
\end{equation}
\end{widetext}
where $\D$ denotes a functional integration measure over all lattice sites; for example
$$
\D\s \equiv \prod_{i=1}^{N}\d\s(t).
$$
The first step is to perform the average over the disorder of expression \refeq{Z} for the generating functional; since the disorder is contained only in the $H_{J}$ part of the hamiltonian, we have to perform the average only on the third line of the \refeq{Z}. Thus we have to compute the integral
\begin{widetext}
\begin{equation}
\begin{split}
\overline{\exp \Bigg[\int dt\ i\left(\bsh(t)\cdot\bm{\partial_{\s}H_{J}}(t)) + \bth(t)\cdot\bm{\partial_{\tau}H_{J}}(t)\right) \Bigg]} =& \\
\pexp{\prod}\int d\pexp{J}\ \exp\Bigg\{-\frac{\pexp{J^{2}}N^{p-1}}{J^{2}p!^2}-&\frac{\pexp{J}}{2^p p!} \int dt\ [i(\sh_{i_1}+\th_{i_1})(\s_{i_2}+\tau_{i_2})\dots(\s_{i_p}+\tau_{i_p})\\
+i(\s_{i_1}+\tau_{i_1})(\sh_{i_2}+\th_{i_2})\dots&(\s_{i_p}+\tau_{i_p}) + \dots + (\s_{i_1}+\tau_{i_1})(\s_{i_2}+\tau_{i_2})\dots i(\sh_{i_p}+\th_{i_p})]\Bigg\},
\end{split}
\end{equation}
where we have symmetrized the $(\sh+\th)_{i_{1}}(\s+\tau)_{i_{2}}\dots(\s+\tau)_{i_{p}}$ couplings; performing the gaussian integration yields
\begin{equation}
\begin{split}
\overline{\exp \Bigg[\int dt\ i\left(\bsh(t)\cdot\bm{\partial_{\s}H_{J}}(t)) + \bth(t)\cdot\bm{\partial_{\tau}H_{J}}(t)\right) \Bigg]}= &\\
\  \frac{J^2}{2^{2p+2}N^{p-1}}\int dt dt' \Big\{p\Big[i(\bsh+\bth)_{t}\cdot &i(\bsh+\bth)_{t'}\Big]\Big[(\bs+\bt)_{t}\cdot(\bs+\bt)\Big]_{t'}^{p-1} + \\
\ p(p-1)\Big[i(\bsh+\bth)_{t}&\cdot(\bs+\bt)_{t'}\Big]\Big[(\bs+\bt)_{t}\cdot i(\bsh+\bth)_{t'}\Big]\Big[(\bs+\bt)_{t}\cdot(\bs+\bt)_{t'}\Big]^{p-2} \Big\}.
\end{split}
\label{averagedZ}
\end{equation}
\end{widetext}
As we anticipated in section \ref{sec:dynamics}, the average over the
disorder has decoupled the lattice sites, at the price of generating a
coupling between configurations of the system at different times.

This dynamical coupling is conceptually similar to the coupling
between replicas that occurs in the static treatment of the $p$-spin
model \cite{Crisanti92,Castellani05}; so, following
\cite{Kirkpatrick87}, we define the dynamical overlaps between
auxiliary MSR fields,
\begin{eqn}
Q_{1}(t,t')  =  \frac{1}{N}\sum_{i=1}^{N}i\sh_{i}(t)i\sh_{i}(t'),\\ 
Q_{2}(t,t')  =  \frac{1}{N}\sum_{i=1}^{N}i\sh_{i}(t)i\th_{i}(t'),\\
Q_{3}(t,t')  =  \frac{1}{N}\sum_{i=1}^{N}i\th_{i}(t)i\sh_{i}(t'),\\ 
Q_{4}(t,t')  =  \frac{1}{N}\sum_{i=1}^{N}i\th_{i}(t)i\th_{i}(t'),
\end{eqn}
between dynamical fields,
\begin{eqn}
Q_{5}(t,t')  =  \frac{1}{N}\sum_{i=1}^{N}\s_{i}(t)\s_{i}(t'),\\
Q_{6}(t,t')  =  \frac{1}{N}\sum_{i=1}^{N}\s_{i}(t)\tau_{i}(t'),\\
Q_{7}(t,t')  =  \frac{1}{N}\sum_{i=1}^{N}\tau_{i}(t)\s_{i}(t'),\\
Q_{8}(t,t')  =  \frac{1}{N}\sum_{i=1}^{N}\tau_{i}(t)\tau_{i}(t'),
\end{eqn}
and between auxiliary and dynamical fields
\begin{eqn}
Q_{9}(t,t')  =  \frac{1}{N}\sum_{i=1}^{N}i\sh_{i}(t)\s_{i}(t') = Q_{13}(t',t)\\ 
Q_{10}(t,t')  =  \frac{1}{N}\sum_{i=1}^{N}i\sh_{i}(t)\tau_{i}(t') = Q_{14}(t',t)\\ 
Q_{11}(t,t')  =  \frac{1}{N}\sum_{i=1}^{N}i\th_{i}(t)\s_{i}(t') = Q_{15}(t',t)\\ 
Q_{12}(t,t')  =  \frac{1}{N}\sum_{i=1}^{N}i\th_{i}(t)\tau_{i}(t') = Q_{16}(t',t)\\ 
\end{eqn}
These functions can be incorporated in the \refeq{averagedZ} by inserting $1$-factors in the form of functional Dirac deltas expressed in exponential form, for example
\begin{equation}
\begin{split}
1 = \int\d Q_{1}\d\l_{1}\exp&\Bigg[\int dtdt' i\l_{1}(t,t')\\
&\times\bigg(NQ_{1}(t,t') - \sum_{i=1}^{N}i\sh_{k}(t)i\sh_{k}(t')\bigg)\Bigg];
\end{split}
\end{equation}
the \refeq{averagedZ} then assumes the form 
\begin{widetext}
\begin{eqn}
& &\int \d^{16}Q\d^{16}\l \exp\left\{ \int dt dt' N \left[\sum_{\mu=1}^{16}i\l_{\mu}Q_{\mu}\right]\right\}\\
& &\times \exp\Bigg\{\frac{NJ^2}{2^{2p+2}}\int dt dt' \Big[p\Big(Q_{1}+Q_{2}+Q_{3}+Q_{4}\Big)\Big(Q_{5}+Q_{6}+Q_{7}+Q_{8}\Big)^{p-1} + \\
& &p(p-1)\Big(Q_{9}+Q_{10}+Q_{11}+Q_{12}\Big)\Big(Q_{13}+Q_{14}+Q_{15}+Q_{16}\Big)\Big(Q_{5}+Q_{6}+Q_{7}+Q_{8}\Big)^{p-2}\Big]\Bigg\} \\
& &\times\exp\left\{-i\int dt dt'\left[\l_{1} i\bsh\cdot i\bsh + \l_{2}i\bsh\cdot i\bth + \dots + \l_{16}\bt\cdot i\bth\right]\right\}.
\end{eqn}
\end{widetext}
This expression contains an exponent proportional to the system size $N$, which in the thermodynamic limit allows us to perform the functional integrals over the $Q$s and $\l$s using the saddle-point method. The saddle-point equations read
\begin{eqnn}
i\l_{1}\dots i\l_{4} &=& -\frac{J^{2}p}{2^{2p+2}}(Q_{5}+Q_{6}+Q_{7}+Q_{8})^{p-1},\\ \nonumber \\
i\l_{5}\dots i\l_{8} &=&-\frac{J^{2}p(p-1)}{2^{2p+2}}(Q_{1}+Q_{2}+Q_{3}+Q_{4})\label{lambda5}\\
& &\times(Q_{5}+Q_{6}+Q_{7}+Q_{8})^{p-2}\nonumber \\
& &-\frac{J^{2}p(p-1)(p-2)}{2^{2p+2}}(Q_{9}+Q_{10}+Q_{11}+Q_{12})\nonumber\\
& &\times(Q_{13}+Q_{14}+Q_{15}+Q_{16})\nonumber  \\
& &(Q_{5}+Q_{6}+Q_{7}+Q_{8})^{p-3},  \nonumber
\end{eqnn} 
\begin{eqnn}
i\l_{9}\dots i\l_{12} &=& -\frac{J^{2}p(p-1)}{2^{2p+2}}(Q_{13}+Q_{14}+Q_{15}+Q_{16})\nonumber\\
&&\times(Q_{5}+Q_{6}+Q_{7}+Q_{8})^{p-2}, \\
i\l_{12} \dots i\l_{16} &=& -\frac{J^{2}p(p-1)}{2^{2p+2}}(Q_{9}+Q_{10}+Q_{11}+Q_{12})\nonumber\\
&&\times(Q_{5}+Q_{6}+Q_{7}+Q_{8})^{p-2}.
\end{eqnn}
The $Q_{1-4}$ can be self-consistently set to zero, as they essentially are the correlation functions for the auxiliary MSR fields; besides this, it is immediate to see that the second term in the \refeq{lambda5} contains the sum of all the response functions for the system $Q_{9-12}$, multiplied for a sum of the same functions with inverted times $Q_{12-16}$; thus, since every response function $R(t,t')$ is zero for $t<t'$ because of causality, the resulting product vanishes and the $\l_{5-9}$ are all equal to zero.

Using the saddle-point equations, we can eliminate the $\l$s and replace the $Q$s in the saddle point with the correlation functions,
\begin{eqn}
Q^{SP}_{5}(t,t') \equiv C_{\s\s}(t,t'),\\
Q^{SP}_{6}(t,t') \equiv C_{\s\tau}(t,t'),\\
Q^{SP}_{7}(t,t') \equiv C_{\tau\s}(t,t'),\\
Q^{SP}_{8}(t,t') \equiv C_{\tau\tau}(t,t'),
\end{eqn}
and the response functions of the system
\begin{eqn}
Q^{SP}_{13}(t,t') \equiv -R_{\s\s}(t,t') = Q^{SP}_{9}(t',t)\\
Q^{SP}_{14}(t,t') \equiv -R_{\s\tau}(t,t') = Q^{SP}_{10}(t',t),\\
Q^{SP}_{15}(t,t') \equiv -R_{\tau\s}(t,t') = Q^{SP}_{11}(t',t)\\
Q^{SP}_{16}(t,t') \equiv -R_{\tau\tau}(t,t') = Q^{SP}_{12}(t',t).
\end{eqn}
If we now define
$$
K_{p} \equiv \frac{J^{2}p}{2^{2p+1}},
$$
and 
\begin{eqn}
\C &=& C_{\s\s}+C_{\s\tau}+C_{\tau\s}+C_{\tau\tau},\\
\R &=& R_{\s\s}+R_{\s\tau}+R_{\tau\s}+R_{\tau\tau},
\end{eqn}
we can rewrite the averaged generating functional in the following way
\begin{widetext}
\begin{equation}
\begin{split}
\overline{Z[0]}=& \int \D\s\D\hat{\s}\D\tau\D\hat{\tau}\ \exp\left[ -\frac{1}{2}\int dt dt'(\bsh D_0\bsh + \bth D_0\bth)\right]\exp\left[-\frac{1}{2}\int dt dt'  K_{p}(\bsh+\bth)\C(t,t')^{p-1}(\bsh+\bth) \right]\\
&\times\exp\int dt\ i\bsh\cdot\Bigg[\dot{\bs} + \mu\bs+\frac{(D-T\log2)}{2}\bt -K_{p}(p-1)\int dt'\ \R(t,t')\C(t,t')^{p-2}(\bs(t')+\bt(t'))\Bigg]\\
&\times\exp\int dt\ i\bth\cdot\Bigg[\dot{\bt} + \nu\bt+\frac{(D-T\log2)}{2}\bs -K_{p}(p-1)\int dt'\ \R(t,t')\C(t,t')^{p-2}(\bs(t')+\bt(t'))\Bigg];\\
\end{split}
\label{myendgen}
\end{equation}
\end{widetext}
as in \cite{Kirkpatrick87}, this functional is now local in space (i.e. refers to single sites only) but nonlocal in time; we can also see that this expression is quite similar to the one we had in equation \refeq{Z}: we have the quadratic terms in $\bsh$ and $\bth$, containing the noise correlators, and the linear ones containing the equations of motion \refeq{myeqsigma}, \refeq{myeqtau} themselves.

\subsection{Equations for the correlation and the response}
Once the effective generating functional has been computed, the self-consistency equations for the correlation and response functions can be readily derived. For the $C_{\s\s}$ we just use the basic definition
\begin{eqn}
\dpart{C_{\s\s}(t_{1},t_{2})}{t_{1}} = \left<\dot{\s}(t_{1})\s(t_{2})\right>,
\end{eqn}
where $\thav{\cdot}$ denotes the average over the gaussian thermal noises $\xi(t)$ and $\zeta(t)$\footnote{The thermal noises $\xi$ and $\zeta$ have a non-diagonal correlation matrix, as it is immediate to see from the \refeq{corr}. However, their probability distribution is still a Gaussian, albeit a non-factorized one. What is important is that the distribution can be factorized in terms where each one is relative to a single site.}; using the \refeq{myeqsigma}, we get
\begin{eqn}
\dpart{C_{\s\s}(t_{1},t_{2})}{t_{1}} &=& -\mu(t_{1})C_{\s\s}(t_{1},t_{2})-\frac{(D-T\log2)}{2}C_{\tau\s}(t_{1},t_{2})\\ 
& &+K_{p}(p-1)\int dt'\ \R(t,t')\C(t,t')^{p-2}\\
& &\times(C_{\s\s}(t',t_{2})+C_{\tau\s}(t',t_{2})) + \left<\x(t_1)\s(t_{2})\right>.
\end{eqn}
The last term has to be computed using directly the general definition of an MSR generating functional (see \cite{Castellani05} for details). First we write
$$
\overline{Z} = z^{N} 
$$
where the $z$ is the reduced generating functional for a single couple of dynamic variables $(\s,\tau)$ only. With this definition we can write $\left<\x(t_1)\s(t_{2})\right>$ as (to lighten the notation, we omit the time contractions).
\begin{equation}
\begin{split}
\left<\x(t_{1})\s(t_{2})\right>& =(-i) \dfunc{ }{j(t_{1})}\Big{|}_{j=0} \int \d\x\ \d\z\ \d\s\ \d\sh\ \d\tau\ \d\th\\
&\times\s(t_{2})\exp\left[-\frac{1}{2}(\x G_{\x\x}\x + 2\x G_{\x\z}\z+\z G_{\z\z}\z)\right]\\ 
&\times\exp\left[i(-\x\sh + \x j -\z\th)\right]\exp\left[\mathcal{L}\right]
\end{split}
\label{}
\end{equation}
where the probability distribution of the noise has been made explicit by defining the inverse of the correlation matrix
$$
G \equiv D^{-1},
$$
\begin{eqn}
D_{\x\x}(t,t') \equiv \thav{\x(t)\x(t')},\\ 
D_{\z\z}(t,t') \equiv \thav{\z(t)\z(t')},\\ 
D_{\x\z}(t,t') \equiv \thav{\x(t)\z(t')}.
\end{eqn}
and all the terms that don't contain the noise fields (the time-derivatives of the dynamic variables, the Lagrange multipliers, the convolutions with the response and correlation functions) have been cropped in $\mathcal{L}$.\\
The functional integral over the noise is a standard Gaussian integral with a linear term, so it can be readily performed, yielding
\begin{equation}
\begin{split}
\int \d\s \d\sh \d\tau \d\th&\ \s(t_{2})\exp\Bigg\{-\frac{1}{2}\big[(-\sh+j)D_{\x\x}(-\sh+j)\\
&+2(-\sh+j)D_{\x\z}(-\th)+\th D_{\z\z}\th\big]\Bigg\}\exp\left[\mathcal{L}\right];
\end{split}
\end{equation}
now, by taking the functional derivative with respect to $j(t_{i})$ and multiplying by $-i$ we get
\begin{equation}
\begin{split}
\left<\x(t_{1})\s(t_{2})\right>= &\int \d\s \d\sh \d\tau \d\th \int dt' \big[D_{\x\x}(t_{1},t')(-i)\s(t_{2})\sh(t')\\
& + D_{\x\z}(t_{1},t')(-i)\s(t_{2})\th(t')\big]e^{S(\s,\tau,\sh,\th)}\\
=&\int dt' \big[D_{\x\x}(t_{1},t')R_{\s\s}(t_{2},t')\\
& + D_{\x\z}(t_{1},t')R_{\s\tau}(t_{2},t')\big],
\end{split}
\end{equation}
where $e^{S(\s,\tau,\sh,\th)}$ is the probability distribution for the dynamic and auxiliary variables induced by the probability distribution for the noise\cite{Martin73}.\\
By using this result, and recalling the definition \refeq{corr} for the correlation matrix, we can finally write the equation for the $C_{\s\s}$
\begin{equation}
\begin{split}
\dpart{C_{\s\s}(t_{1},t_{2})}{t_{1}} = &-\mu(t_{1})C_{\s\s}(t_{1},t_{2})\\
&-\frac{(D-T\log2)}{2}C_{\tau\s}(t_{1},t_{2})\\
&+K_{p}(p-1)\int_{-\infty}^{t_{1}} dt'\ \R(t,t')\C(t,t')^{p-2}\\
&\times(C_{\s\s}(t',t_{2})+C_{\tau\s}(t',t_{2}))\\
&+2TR_{\s\s}(t_{2},t_{1}) +K_{p}\int_{-\infty}^{t_{2}} dt' \C(t,t')^{p-1}\\
&\times(R_{\s\s}(t_{2},t')+R_{\s\tau}(t_{2},t'));
\end{split}
\end{equation}
the term proportional to the $R_{\s\s}$ is zero since we assume $t_{1}>t_{2}$. The same method can be applied to the remaining correlation functions.

For what concerns response functions, it can be easily proven that\cite{Castellani05,Bouchaud96}
\begin{equation}
\thav{\dfunc{\s(t_{1})}{\xi(t_{2})}} = \dfunc{\thav{\s(t_{1})}}{h(t_{2})},
\label{rssdef}
\end{equation}
which in turn implies
\begin{eqn}
\dpart{R_{\s\s}(t_{1},t_{2})}{t_{1}} =\dpart{}{t_{1}}\thav{\dfunc{\s(t_{1})}{\xi(t_{2})}}= \thav{\dfunc{\dot{\s}(t_{1})}{\xi(t_{2})}},
\end{eqn}
and so
\begin{equation}
\begin{split}
\dpart{R_{\s\s}(t_{1},t_{2})}{t_{1}} =& -\mu(t_{1})\thav{\dfunc{\s(t_{1})}{\xi(t_{2})}}\\
& -\frac{(D-T\log2)}{2}\thav{\dfunc{\tau(t_{1})}{\xi(t_{2})}}\\
&+K_{p}(p-1)\int_{-\infty}^{t_{1}} dt'\ \R(t,t')\C(t,t')^{p-2}\\
&\times\left[\thav{\dfunc{\s(t')}{\xi(t_{2})}} + \thav{\dfunc{\tau(t')}{\xi(t_{2})}}\right]+\thav{\dfunc{\xi(t_{1})}{\xi(t_{2})}},
\label{eqrss}
\end{split}
\end{equation}
The other equations can be obtained by trivially generalizing the \refeq{rssdef}.\\
Let us make an important remark. The last term in the \refeq{eqrss} (given by the functional derivative of the noise with respect to itself) is a Dirac delta, which assures that
$$
\lim_{t_{1}\to t_{2}^{+}}R_{\s\s}(t_{1},t_{2}) = 1.
$$
The same applies to $R_{\tau\tau}$. On the other hand, the equation for the $R_{\s\tau}$ is the following
\begin{equation}
\begin{split}
\dpart{R_{\s\tau}(t_{1},t_{2})}{t_{1}} =& -\mu(t_{1})R_{\s\tau}(t_{1},t_{2}) -\frac{(D-T\log2)}{2}R_{\s\s}(t_{1},t_{2})\\
&+K_{p}(p-1)\int_{-\infty}^{t_{1}} dt'\ \R(t,t')\C(t,t')^{p-2}\\
&\left(R_{\s\tau}(t',t_{2}) + R_{\s\s}(t',t_{2})\right),
\label{eqrst}
\end{split}
\end{equation}
which indeed lacks the Dirac delta, implying
$$
\lim_{t_{1}\to t_{2}^{+}}R_{\s\tau}(t_{1},t_{2}) = 0,
$$
which applies to the $R_{\tau\s}$ too. This relation is fundamental to recover the static equation for the density (see appendix \ref{app:dens}).

The last thing left is computing the expression for the Lagrange multipliers $\mu(t)$ and $\nu(t)$; since we introduced those terms to enforce the spherical constraints, their expression must be determined self-consistently from them,
\begin{eqn}
C_{\s\s}(t,t) = N &\Rightarrow& \frac{dC_{\s\s}(t,t)}{dt} = 0,\\
C_{\tau\tau}(t,t) = N &\Rightarrow& \frac{dC_{\tau\tau}(t,t)}{dt} = 0.\
\end{eqn}
By expressing these total derivatives in terms of the dynamical equations for the correlation functions, we get, for the $\mu(t)$
\begin{widetext}
\begin{equation}
\begin{split}
\mu(t) =&-\frac{(D-T\log 2)}{2}(2n(t)-1)\\
&+K_{p}(p-1)\int_{-\infty}^{t}dt' \R(t,t')\C(t,t')^{p-2}(C_{\s\s}(t',t)+C_{\tau\s}(t',t)) \\
&+TR_{\tau\tau}(t,t)+K_{p}\int_{-\infty}^{t}dt' \C(t,t')^{p-1}(R_{\s\s}(t,t')+R_{\s\tau}(t,t')) = 0,
\label{eqmu}
\end{split}
\end{equation}
and for the $\nu(t)$,
\begin{equation}
\begin{split}
\nu(t) = & -\frac{(D-T\log 2)}{2}(2n(t)-1)\\
&+K_{p}(p-1)\int_{-\infty}^{t}dt' \R(t,t')\C(t,t')^{p-2}(C_{\tau\tau}(t',t)+C_{\s\tau}(t',t))\\
&+TR_{\s\s}(t,t)+K_{p}\int_{-\infty}^{t}dt' \C(t,t')^{p-1}(R_{\tau\tau}(t,t')+R_{\tau\s}(t,t')) = 0.
\label{eqnu}
\end{split}
\end{equation}
\end{widetext}
These are the expressions for $\mu(t)$ and $\nu(t)$ that are to be plugged in the equations for the correlation and response functions.

From this point, the derivation proceeds by imposing the FDT to couple the response and correlation functions, and then using the TTI to manipulate the resulting expressions. Those steps are only a matter of standard algebra, so we shall not report them.

\subsection{The $d\to1$ limit \label{sec:dto1}}
The Blume Capel p-spin spherical model is supposed to be indistinguishable for a plain PSM in the $d\to1$ limit, since in this limit the sites are all filled-in. So, our equations are supposed to give back the classic dynamical equation for the PSM when $d\to1$.

From equation \refeq{A}, we can see that $C_{diff}(t)$ is automatically zero for all times when d=1. So we only need to take care of equation \refeq{eqdinamica}, but the problem here is more tricky. In fact, it is immediate to verify that by just sending $d$ to one in the \refeq{eqdinamica}, we get a wrong result.\\
The problem is that the value of $d$ depends strongly on the point (D,T) of the phase plane by means of equation \refeq{eqstatic}; indeed, if we invert the picture and we treat $d$ as a free parameter, we see that when $d\to1$, then the absolute value of the crystal field $D$ tends to infinity. This is perfectly reasonable; when $D$ becomes lower and lower, the configurations with a good number of empty sites are energetically disadvantaged, and the partition function is dominated mainly by configurations with many filled-in sites, which become the only ones accessible to the system when $D\to -\infty$.
Since $|D|\to\infty$ at the same time as $d$ approaches one, the latter limit is nontrivial and has to be taken carefully.\\
Therefore, in order to estimate the order of magnitude of $D$ as $d$ tends to one, we solve the \refeq{eqstatic} for $D$, obtaining
$$
D = \frac{3 d^{p+1}-3 d^p+d^2 T^2 \log16 -4 T^2 d (\log2-1)-2 T^2}{4 (d-1) Td}.
$$
By plugging this into equation \refeq{eqdinamica}, we have
\begin{eqn}
\dot{C}(t) &=& \left[T - \frac{T}{2d} - T\right]C(t) \nonumber\\
& &-\frac{J^{2}p(d)^{p-1}}{4T} \int_{0}^{t}du\ C^{p-1}(t-u)\dot{C}(u);
\end{eqn}
by imposing $d=1$, this becomes
$$
\dot{C}(t) = - \frac{T}{2}C(t) -\frac{J^{2}p}{4T} \int_{0}^{t}du\ C^{p-1}(t-u)\dot{C}(u),
$$
which is the dynamical equation for the PSM \cite{Crisanti93,Kirkpatrick87,Castellani05}, apart from a factor $1/2$ in front of the r.h.s.. Its presence is due to the fact that in the limit $d\to1$, $\s = \tau$ and so our model contains $2N$ $\s$ spins, while the variance of the $J$ couplings has the usual scaling for a system of $N$ spins.
By taking the long time limit, we get
\begin{equation}
\frac{q}{1-q} = \frac{p\beta^{2}}{2}q^{p-1} \label{eq:qdyn}
\end{equation}
which yields the well-established values for the dynamic transition temperature $T_{d}$ and the dynamical overlap $q_{d}$
\begin{eqnn}
q_{d} = \frac{p-2}{p-1} &\qquad&T_{d} = \sqrt{\frac{p(p-2)^{p-2}}{2(p-1)^{p-1}}}. 
\end{eqnn}
This shows that, since in the long time limit $\dot{C}(t)$ is always zero, the $1/2$ factor is completely irrelevant for what concerns the properties of the system at the transition point.

\subsection{The equation for the density \label{app:dens}}

In this work, we have stressed multiple times the fact that the static equation \refeq{eqstatic} is needed to solve the dynamics of the system, since the equilibrium value of $d$ is not known \emph{a priori}. This is fine for all practical purposes, but clearly unsatisfying from the theoretical point of view, as the \refeq{eqstatic} has been derived using replica theory in a static scenario. It is then necessary to show that equation \refeq{eqstatic} can be derived also in the MSR formalism in order to get a completely auto-consistent theoretical picture.

\begin{figure}[htb!]
\begin{center}
\includegraphics[width = 0.99\columnwidth]{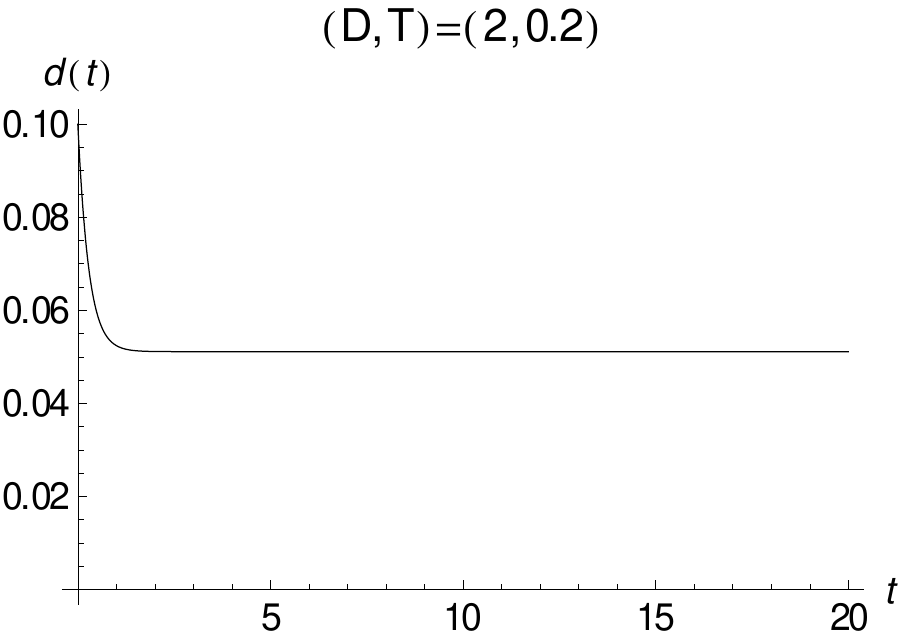}
\includegraphics[width = 0.99\columnwidth]{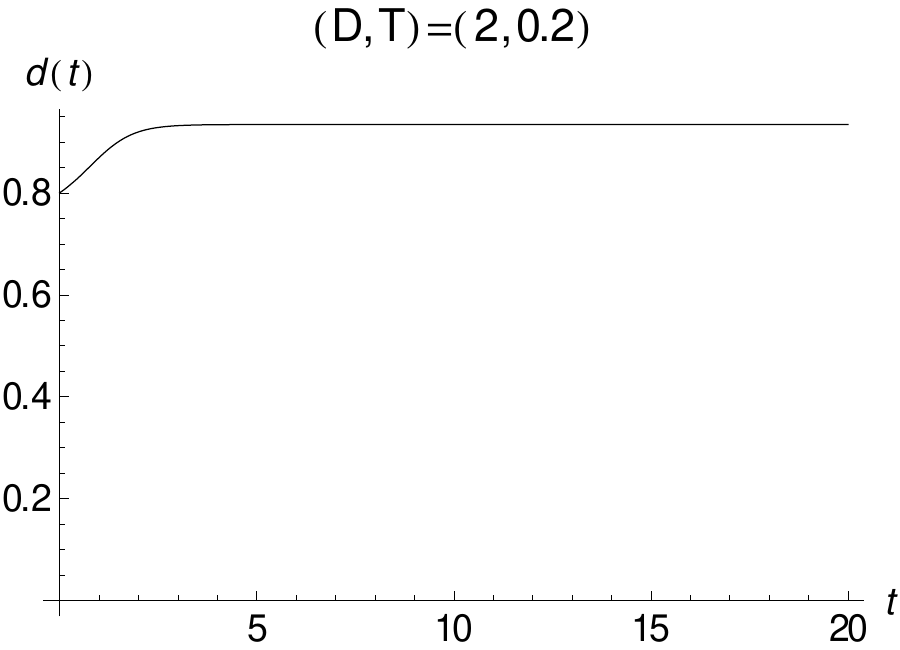}
\caption{Solutions of the \refeq{eqstatic-dynamic} for $(D,T) = (0.2,2)$, with $d(0)=0.1$ (up), e $d(0)$ = 0.8 (down); both solutions converge rapidly to their respective equilibrium values, $d^{-} = 0.051100$ and $d^{+} = 0.934117$.}
\label{n(t)1}
\end{center}
\end{figure}

\begin{figure}[htb!]
\begin{center}
\includegraphics[width = 0.99\columnwidth]{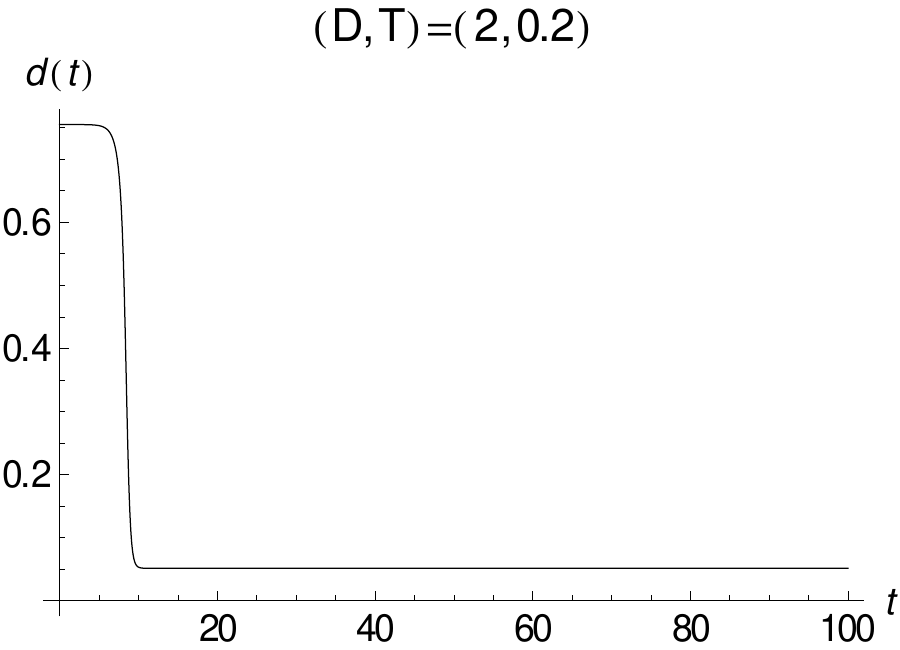}
\includegraphics[width = 0.99\columnwidth]{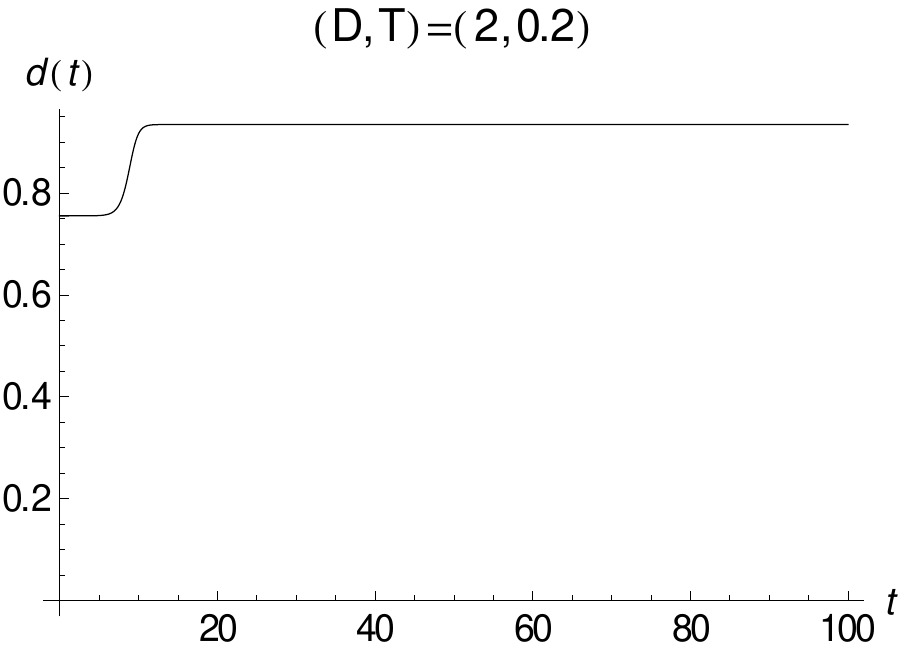}
\caption{Solutions of the \refeq{eqstatic-dynamic} for $(D,T) = (0.2,2)$, with $d(0)=0.754865$. slightly below $d_{int}$ (up), and $d(0) = 0.754867$ slightly above $d_{int}$ (down). $d_{int} =0.754866$. The solution ends up in a different attractor depending on the choice of the initial value $d(0)$.}
\label{n(t)2}
\end{center}
\end{figure}

Our starting point is a trivial rephrasing of relation \refeq{trasf}
$$
n_{i}(t) = \frac{\s_{i}(t)\tau_{i}(t)+1}{2}.
$$
We sum this relation over lattice sites, divide by the system size $N$ and then take the thermal average on both sides. We get the following relation between the density and the $\s\tau$ correlation function
\begin{equation}
\thav{d(t)} = \frac{C_{\s\tau}(t,t)+1}{2}.
\end{equation}
By taking the total derivative with respect to time we get
\begin{equation}
\begin{split}
\dtot{\thav{d(t)}}{t} = &\frac{1}{2}\dtot{C(t,t)}{t}\\ 
= &\frac{1}{2}\left(\dpart{C_{\s\tau}(t_{1},t_{2})}{t_{1}} + \dpart{C_{\s\tau}(t_{1},t_{2})}{t_{2}} \right)\Bigg|_{t_{1}=t_{2}=t},
\end{split}
\end{equation}
which can be then rewritten in terms of the dynamical equations for the $C_{\s\tau}$
\begin{equation}
\begin{split}
\dtot{\thav{d(t)}}{t} &=\ -\mu(t)(2\thav{d(t)}-1) - \left(\frac{D-T\log 2}{2}\right)\\ 
&+ \frac{J^{2}p(p-1)}{2^{2p+1}}\int_{-\infty}^{t} dt'\ \R(t,t')\C(t,t')^{p-2}\\
&\times(C_{\s\tau}(t',t)+C_{\tau\tau}(t',t))+2TR_{\tau\s}(t,t)\\ 
&+ \frac{J^{2}p}{2^{2p+1}}\int_{-\infty}^{t} dt\ \C^{p-1}(t,t')(R_{\tau\s}(t,t')+ R_{\tau\tau}(t,t')).
\label{eqdensityinitial}
\end{split}
\end{equation}
Now, given that $\lim_{t_{1}\to t_{2}^{+}}R_{\s\tau}(t_{1},t_{2}) = 0$, and using the TTI and FDT, we get
\begin{equation}
\begin{split}
\dtot{\thav{d(t)}}{t} = &-\mu(t)(2d(t)-1) - \left(\frac{D-T\log 2}{2}\right)\\ 
&-\frac{J^{2}p^{2}}{2^{2p+2}T}\int_{0}^{\infty}\C(u)^{p-1}\dot{\C}(u)du,
\end{split}
\end{equation}
where we have switched to the notation $\thav{d(t)} = d(t)$.\\
The expression \refeq{eqmu} for the Lagrange multiplier can be manipulated in the same way, obtaining
\begin{equation}
\begin{split}
\mu(t) = &-\frac{D-T\log2}{2}(2d(t)-1)\\
&-\frac{K_{p}}{2T}\int_{0}^{\infty}du\ \dot{\C}(u)\C(u)^{p-1} + T;
\end{split}
\end{equation}
note that, since we are looking for a dynamical equation, we are now working with a time-dependent $d$.\\
We can now see that the integrals in both those expressions can be easily computed using the substitution formula. By doing so and than plugging the expression for $\mu$ in the equation we get
\begin{equation}
\begin{split}
\dot{d}(t) =\ &(2d(t)-1)^{2}\left(\frac{D-T\log2}{2}\right) - \left(\frac{D-T\log2}{2}\right)\\
&-\frac{J^{2}p}{2^{2p+2}T}(4d(t))^{p}(2d(t)-1)\\
& + \frac{J^{2}p}{2^{2p+2}T}(4d(t))^{p} -T(2d(t)-1),
\end{split}
\end{equation}
which, after some trivial algebra, leads to
\begin{equation}
\dot{d} = \ 2d(d-1)(D-T\log2)-\frac{p}{2T}d^{p}(d-1) - T(2d-1).
\label{eqstatic-dynamic}
\end{equation}
If we now assume the system to be at equilibrium at all times
$$
\dot{d} = 0 \qquad d(t) = d_{eq}
$$
the \refeq{eqstatic-dynamic} can be rewritten as
$$
\frac{p}{2}d^{p}(1-d) = T^{2}(2d-1) + 2(TD-T^{2}\log 2)d(1-d),
$$
which indeed is the static equation \refeq{eqstatic} for the density. This is the result we wanted.\\
At this point, the reader has probably noticed the fact that, in this derivation, we have used the TTI and FDT, which are suited to an \emph{equilibrium} situation, to derive a dynamical equation for the thermal average of a one-time observable, which is supposed to vary only in a \emph{nonequilibrium} scenario.\\
This indeed is not a problem in itself. We could have embraced the equilibrium picture from the very beginning (i.e., setting $\dtot{C(t,t)}{t}=0$ in the first place), and we would have anyway recovered the static equation for the density, which was the only result we needed for our dynamical study to be self-contained. However, the dynamic equation \refeq{eqstatic-dynamic} can be interesting to study in itself; by completing it with an initial value for the density $d(0)$, we obtain a Cauchy problem for which an unique solution is guaranteed to exist. We plot four such solutions in figures \ref{n(t)1} and \ref{n(t)2} for (D,T) =(2,0.2).

The plots show that, in the phase coexistence region, both equilibrium
solutions correspond to an attractor for the dynamics, which is not
surprising; where the solution actually ends up depends on the initial
value d(0) that is chosen.\\ Besides this, it is interesting to notice
that the unstable root $d_{int}$ of equation \refeq{eqstatic} (which,
as we said in section \ref{sec:statics}, is always to be discarded on
grounds of replica stability arguments) is not an attractor for the
dynamics, and actually is the critical value of d(0) that marks the
separation point between the two attractors. Thus, the selection rule
of section \ref{sec:statics} for the value of the density is recovered
from a dynamical point of view.

We conclude this section with a final remark. It may seem contradictory that the density $d(t)$ evolves in time, since it is a one-time quantity which is assumed to be constant in an equilibrium dynamics scenario. However, this is true only if the initial value $d(0)$ (which is completely arbitrary) is chosen different from both of the two equilibrium solutions of the static equation, which is exactly the contrary of what condition \refeq{densitycondition} requires.\\
Thus, the \refeq{densitycondition} can be seen both as a reasonable physical assumption, and as a self-consistency condition for the equilibrium dynamics.

\bibliography{Lucabib_221014}

\end{document}